\documentclass[aps,prl,article,twocolumn,preprintnumbers,amsmath,amssymb,superscriptaddress,longbibliography]{revtex4-2}

\usepackage{graphicx}  
\usepackage{dcolumn}   
\usepackage{bm}        
\usepackage{amssymb}   
\usepackage{amsmath}
\usepackage{mathrsfs}
\usepackage{physics}
\usepackage{epigraph}
\usepackage{braket}
\usepackage{gensymb}
\usepackage{lmodern}
\usepackage{ tipa }
\usepackage{bbold}
\usepackage{esint}
\usepackage{mathdots}
\usepackage{xcolor}
\usepackage{appendix}
\usepackage{natbib}
\usepackage{multirow}
\usepackage{float}
\usepackage{xr}
\usepackage[normalem]{ulem}
\usepackage{times}
\usepackage{comment}
\usepackage[hidelinks,colorlinks=true, linkcolor=blue, urlcolor=blue, citecolor=blue]{hyperref}

\newcommand{\Ham}{\mathcal{H}}

\newcommand{\ii}{\mathrm{i}}

\newcommand{\SISec}{Supplementary Note}

\usepackage{xcolor}

\colorlet{BLACK}{black}\colorlet{RED}{red}
\begin{document}
\title{Transport-Noise and Optical Witnesses of Electronic Multipartite Entanglement} 	

\date{\today}
\author{Shuhan Ding}\thanks{These authors contributed equally to this work.}
\affiliation{Department of Chemistry, Emory University, Atlanta, GA 30322, USA}
\author{Prakash Sharma}\thanks{These authors contributed equally to this work.}
\affiliation{Department of Chemistry, Emory University, Atlanta, GA 30322, USA}
\author{Zecheng Shen}
\affiliation{Department of Chemistry, Emory University, Atlanta, GA 30322, USA}
\author{Jiang-Xiazi Lin}
\affiliation{Department of Physics, Emory University, Atlanta, GA 30322, USA}
\author{Sergei Urazhdin}
\affiliation{Department of Physics, Emory University, Atlanta, GA 30322, USA}
\author{Yao Wang}
\email[Correspondence should be addressed to \href{mailto:yao.wang@emory.edu}{yao.wang@emory.edu}]{}
\affiliation{Department of Chemistry, Emory University, Atlanta, GA 30322, USA}
\date{\today}
\begin{abstract} 
Entanglement among particles distinguishes correlated quantum materials from conventional metals and semiconductors. Experimentally certifying intrinsic entanglement among interacting electrons in materials is important not only for classifying states of matter, but also for developing material-based quantum technologies. Here, we introduce a transport protocol for witnessing multipartite entanglement from the equilibrium current-noise spectrum, measured directly or reconstructed from optical conductivity. The appropriately integrated, symmetrized, and projected noise spectrum obeys an upper bound derived from microscopic model parameters. Unlike witnesses based on localized spins, our framework applies to general electronic states without requiring magnetic moments and remains invariant under single-particle basis transformations. We benchmark it in twisted bilayer graphene, twisted bilayer MoTe$_2$, and Hubbard models, and demonstrate entanglement certification in a fractional Chern insulator. This approach extends recently developed scattering-based entanglement witnesses to ultralow-temperature conditions, where many candidate entangled phases emerge.
\end{abstract}

\maketitle

\section{Introduction}
Quantum materials are distinguished by collective electronic behavior that cannot be captured by traditional single-particle descriptions\,\cite{keimer2017physics, basov2017towards}. Over the past decades, this has led to an expanding landscape of exotic quantum phases. Representative examples include quantum spin liquids\,\cite{balents2010spin}, fractional quantum Hall states\,\cite{bernevig2025fractional,bergholtz2013topological}, and the pseudogap and strange-metal phases\,\cite{greene2020strange,keimer2015quantum,hufner2008two}. These phases often involve strong quantum fluctuations and inter-particle entanglement beyond a quasiparticle description. Establishing reliable ways to certify and quantify entanglement in such materials is therefore important not only for elucidating their fundamental physics, but also for advancing quantum technologies.

Bell inequality violations provide a well-established route to detect entanglement in quantum optics\,\cite{aspect1982experimental}. Complementary approaches in ultracold atoms include spin squeezing\,\cite{esteve2008squeezing} and many-body interferometry\,\cite{islam2015measuring}. Extending such detection to macroscopic materials, however, is substantially more challenging. Recent advances have nevertheless translated these concepts to materials, where correlation-based quantities, including quantum Fisher information (QFI)\,\cite{pezze2009entanglement, hyllus2012fisher, toth2012multipartite, hauke2016measuring, menon2023multipartite, fang2025amplified, shen2026witnessing}, two-tangle\,\cite{coffman2000distributed}, and the two-particle reduced density matrix (2CRDM)\,\cite{liu2025entanglement}, can provide indirect signatures of entanglement. These metrics can be extracted from materials' spectroscopies and have successfully witnessed spin entanglement in magnetic materials\,\cite{mathew2020experimental, scheie2021witnessing, laurell2021quantifying, scheie2024proximate, mazza2026quantum}. However, they rely on high-resolution neutron or x-ray scattering and therefore impose stringent requirements on sample size, environmental control, and ultrahigh spectral resolution. These requirements exclude many candidate platforms such as flat-band two-dimensional materials at sub-kelvin temperatures. This is especially constraining because these systems manifest signatures of quantum correlations such as fractionalized excitations\,\cite{depicciotto1997direct, dolev2008observation}. Yet a direct certification of entanglement in these materials comparable to a Bell-type test remains out of reach.

\begin{figure}[!b]
    \centering\vspace{-4mm}
    \includegraphics[width=\linewidth]{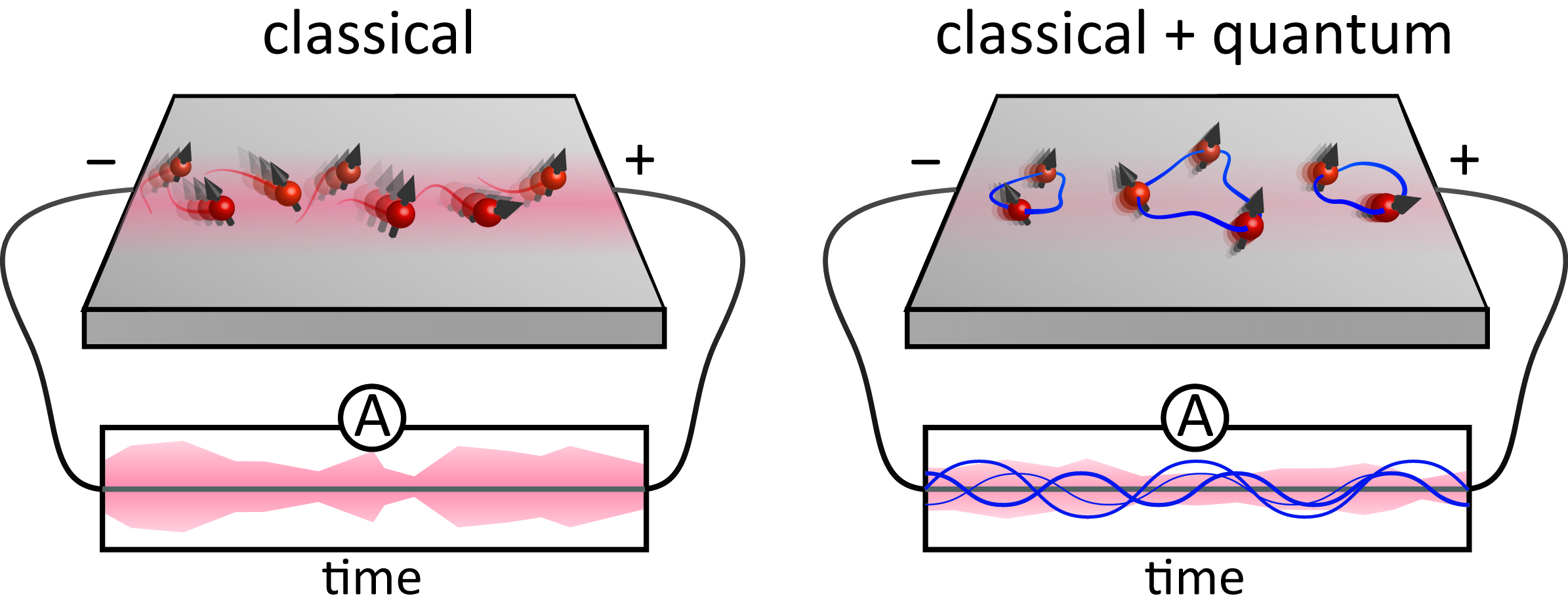}\vspace{-4mm}
    \caption{\textbf{Schematic illustrating current noise from classical and quantum sources.} Left: Thermal current fluctuations in a classical system, where the low-frequency Johnson noise is set by the thermal distribution of electronic velocities. Right: Quantum fluctuations in a quantum system, whose two-time current correlation contains particular dynamical structures generated by entanglement. }
    \label{fig:cartoon}
\end{figure}

To this end, we introduce a current-noise protocol for witnessing entanglement. As illustrated in Fig.~\ref{fig:cartoon}, while noise is present in all transport measurements, its classical and quantum components have fundamentally different manifestations. Classical fluctuations follow thermal statistics and are captured by the standard Johnson noise; quantum fluctuations, by contrast, carry dynamical structures that encode intrinsic many-body electronic correlations. These fluctuations are reflected in the dynamical structure of the noise spectrum, accessible by transport measurements at low frequencies and terahertz and optical spectroscopies at high frequencies. By extracting the quantum current fluctuations from noise spectroscopy, we identify entanglement through the violation of sum-rule bounds determined by intrinsic material properties and the entanglement depth. Symmetry constraints and band projection further tighten these multipartite-entanglement bounds. We benchmark this framework using a variety of experimentally relevant models, establishing a scalable and cryo-compatible route for certifying entanglement in quantum materials. 

\section{Results}
\subsection{Current-noise-based entanglement witness}\label{sec:theory_noise_qfi_witness}
\vspace{-3mm}

We begin by formulating an entanglement witness for interacting electrons based on current fluctuations. This setting differs fundamentally from spin or qubit systems, where the tensor-product structure is uniquely fixed by the underlying modes\,\cite{hyllus2012fisher}. In electronic systems, by contrast, the fermionic antisymmetrization forbids the tensor-product structure and the local single-particle basis remains subject to unitary rotations\,\cite{ghirardi2002entanglement, ghirardi2004general, amico2008entanglement}. Specifically, in a given basis where creation operators are ${c_m^\dagger}$, a pure $k$-producible state is constructed by partitioning the modes into disjoint sets $\mathcal{M}_\ell$\,\cite{almeida2021from, liu2025entanglement}
\begin{equation}\label{eq:kProducibleState}
    \ket{\Psi_{k\text{-prod}}} =  e^{-i\sum_{ij} c_i^\dagger \xi_{ij} c_j} \prod_{\ell} \hat{ \mathcal{C}}_\ell^\dagger \ket{0}\,,
\end{equation}
where $\xi$ is a Hermitian matrix accounting for the basis ambiguity. Each block creation operator is $\hat{\mathcal{C}}_\ell^\dagger=\sum_{\eta_\ell} \phi_\ell^\star(\eta_\ell) \prod_{m \in \mathcal{M}_\ell} (c_m^\dagger)^{\eta_\ell(m)}$, creating at most $k$ fermions. Here, $\eta_\ell(m)$ is binary and $\phi_\ell^\star(\eta_\ell)$s are the irreducible coefficients associated with the block. This definition extends naturally to mixed states via convex mixtures\,\cite{liu2025entanglement}.

In QFI theory, multipartite entanglement is certified through fluctuations of a Hermitian observable that is local in the selected mode basis (see \SISec\ I). Within transport measurements, an experimentally accessible  observable is the symmetrized current-noise spectrum\,\cite{blanter2000shot,clerk2010introduction},
\begin{equation}\label{eq:symmetrizedNoiseSpec}
    S_{\alpha}(\omega, T) = \int_{-\infty}^{\infty}dt\,e^{\ii\omega t}\, \frac{1}{2} \left\langle \left\{ \delta\hat J_{\alpha}(t),\ \delta\hat J_{\alpha}(0) \right\} \right\rangle \,,
\end{equation}
where $\delta\hat J_\alpha=\hat J_\alpha-\langle\hat J_\alpha\rangle$ is the current fluctuation along the direction $\alpha=x,y$. Relevant for transport, we consider the long-wavelength limit, for which $\hat J_{\alpha}=-\partial\Ham/\partial A_\alpha$ with $\Ham$ denoting the Hamiltonian and $A_\alpha$ being the vector potential in the temporal gauge. Here, we focus on the zero-voltage-bias ($A_\alpha\rightarrow 0$) limit. Experimentally, $S_{\alpha}(\omega, T)$ can be directly accessed through equilibrium current noise measurements\,\cite{schoelkopf1997frequency, zakka2007experimental}. At higher frequencies, $S_{\alpha}(\omega,T)$ can be indirectly reconstructed from terahertz and optical measurements using the fluctuation--dissipation theorem\,\cite{kubo1957statistical} (see \SISec\ II):
\begin{equation}\label{eq:noiseOpticalConductivity}
    S_{\alpha}(\omega,T) = \mathcal A\hbar\omega\coth\left(\frac{\hbar\omega}{2k_B T}\right)\sigma_{1,\alpha\alpha}(\omega,T)\,,
\end{equation}
where $\mathcal{A}$ denotes the area of the material and $\sigma_{1,\alpha\alpha}(\omega,T)$ denotes the real-part optical conductivity. 

The measured noise spectrum can be converted into the QFI of $\hat J_{\alpha}$ through\,\cite{hauke2016measuring}
\begin{equation}\label{eq:qfiNoiseSpectrum}
    F_Q[\hat J_\alpha](T) = \frac{4}{\pi} \int_{0}^{\infty}d\omega\, \tanh^2\left(\frac{\hbar \omega}{2k_B T}\right) S_{\alpha}(\omega,T)
\end{equation}
The $\tanh$ kernel removes the purely classical contribution and isolates the quantum part of the current noise.

Unlike prior work that related quantum noise to bipartite entanglement entropy in non-interacting fermions\,\cite{klich2009quantum}, our framework targets multipartite electronic entanglement in interacting systems rather than reconstructing the entropy of a spatial partition. Analogous to Bell test, the idea of entanglement witness is to determine the tightest possible upper bound on the measurable $F_Q$ over all $k$-producible states. Unlike spin operators, the operator $\hat J_{\alpha}=\sum_{ij}[J_\alpha]_{ij}c_i^\dagger c_j$ carries intrinsically mode-dependent coefficients $[J_\alpha]_{ij}$. Its QFI therefore does not admit a universal bound of the form similar to spin systems\,\cite{hyllus2012fisher, hauke2016measuring}. Instead, the appropriate bound is formally the supremum over the full manifold of $k$-producible wavefunctions:
\begin{equation}\label{eq:QFIWitnessInequality}
    F_Q[\hat J_{\alpha}](T) \le \mathcal{A}\cdot\mu(k)=\sup\mkern-2mu\left\{F_Q[\hat J_{\alpha}]\mkern-2mu: \forall\ket{\Psi_{k-\textrm{prod}}}\right\} \,.
\end{equation}
To obtain an explicit supremum, we diagonalize $[J_\alpha]_{ij}$ and sort its eigenvalues in descending/ascending order, denoted as $\{v_{i}^{(\downarrow/\uparrow)}\}_{i=1}^{N}$. In the $k\ll N$ limit, this leads to the explicit supremum
\begin{equation}\label{eq:QFIBoundRes}
    \mu(k)= \frac{k}{\mathcal{A}} \sum_{i=1}^{N_{\rm p}} \left( v_i^{(\uparrow)} - v^{(\downarrow)}_i \right)^2 +\mathcal{O}\left(\frac{kN_{\rm p}}{\mathcal{A}^2}\right)\,.
\end{equation}
Here, $N_{\rm p}$ represents the smaller of the particle and hole numbers. Physically, this supremum is obtained by maximizing the current fluctuations within each $k$-particle block through GHZ-like superpositions of configurations carrying opposite currents (see \SISec\ III for the proof).  

\begin{figure*}[!t]
    \hspace*{-4mm}  \centering
    \includegraphics[width=183mm]{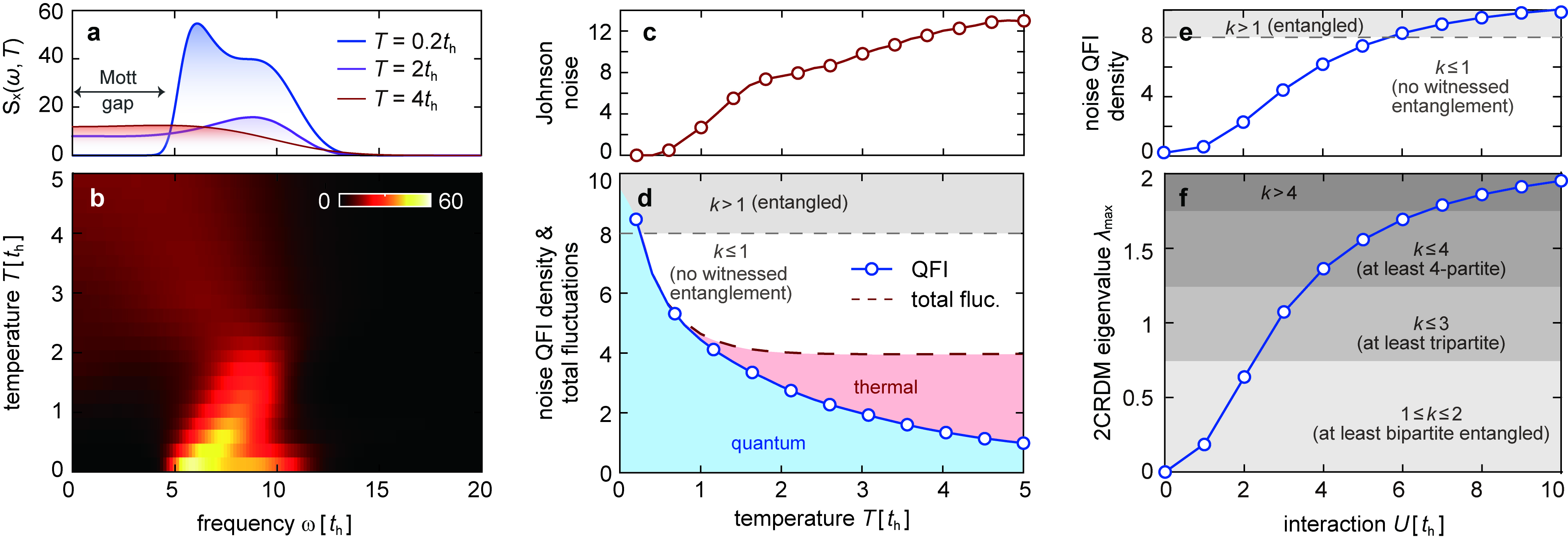}\vspace{-2mm}
    \caption{\textbf{Noise spectrum and entanglement witness in the Hubbard model.} \textbf{a}, Symmetrized current-noise spectrum $S(\omega,T)$ of the half-filled 1D Hubbard model, calculated by DQMC for three representative temperatures: $T=0.2t_h$ (blue), $2t_h$ (purple), and $4t_h$ (red). \textbf{b}, Full temperature evolution of the noise spectrum shown as a false-color map. \textbf{c}, Johnson noise extracted from the $\omega=0$ spectral intensity in \textbf{b}. \textbf{d}, Temperature dependence of the current-noise QFI density $F_Q[\hat J_x]/\mathcal{A}$ (dots and solid line) together with the normalized total current fluctuation $4\langle\delta\hat J_{\alpha}^2\rangle/\mathcal{A}$ (dashed), with their difference representing the thermal fluctuation. \textbf{e}, \textbf{f}, Zero-temperature interaction dependence of the current-noise QFI (\textbf{e}) and the maximal eigenvalue of the 2CRDM (\textbf{f}), obtained using DMRG. The shaded regions in \textbf{d}, \textbf{e}, and \textbf{f} indicate the corresponding thresholds for $k$-producible states. The unit of current-noise QFI in \textbf{d} and \textbf{e} is $(et_h/\hbar)^2$.}
    \label{fig:HubbardBenchmark}
\end{figure*}

Although Eq.~\eqref{eq:QFIBoundRes} is more complicated than the constant-weight bounds used for spin or fermionic-mode QFI\,\cite{hyllus2012fisher, almeida2021from}, this complexity is essential for witnessing electronic entanglement with transport and optical measurements. The supremum $\mu(k)$ obtained after the diagonalization yields a basis-invariant upper bound that naturally incorporates the basis ambiguity in Eq.~\eqref{eq:kProducibleState}. In this construction, the non-interacting Fermi sea is identified as separable and thus provides the proper reference for noise spectral measurements. Consequently, an observed current-noise QFI density satisfying $F_Q/\mathcal{A}>\mu(k)$ excludes all $k$-producible states and certifies a basis-invariant entanglement depth of at least $k+1$ (see \SISec\ III). 

\subsection{Noise spectrum in a Hubbard model}\label{sec:hubbard_benchmark}

To illustrate how entanglement can be certified from the noise spectrum, we consider the single-band Hubbard model as a representative platform (see \textbf{Methods} for the Hamiltonian). This model describes correlated electrons in cuprates and Mott insulators. Its properties can be simulated in a controlled manner using determinant quantum Monte Carlo (DQMC) for finite temperatures and density-matrix renormalization group (DMRG) for zero temperature.

We begin by simulating the temperature dependence of the symmetrized current-noise spectrum $S_{x}(\omega, T)$ defined in Eq.~\eqref{eq:symmetrizedNoiseSpec} using DQMC. As shown in Figs.~\ref{fig:HubbardBenchmark}\textbf{a,b}, for the representative case $U=8t_h$, the spectral weight generally grows as the temperature is lowered, even though thermal fluctuations are expected to weaken. This trend is in sharp contrast with the Johnson noise at $\omega=0$\,\cite{johnson1928thermal,nyquist1928thermal}. In particular, the broad spectral weight below the Mott gap ($\omega\sim 4t_h$) shifts progressively toward higher frequencies as the temperature decreases, depleting the low-frequency noise (see Fig.~\ref{fig:HubbardBenchmark}\textbf{c}). At the same time, high-frequency fluctuations across the Mott gap grow rapidly, signaling the increasing dominance of quantum fluctuations in the low-temperature regime.

The combined contribution of both low- and high-frequency fluctuations can be captured by integrating the full noise spectrum, i.e., $4\langle\delta\hat J_{\alpha}^2\rangle (T) =\int_{0}^{+\infty} d\omega\, 4S_{\alpha}(\omega, T) /\pi $. As shown by the red dashed curve in Fig.~\ref{fig:HubbardBenchmark}\textbf{d}, the resulting total current fluctuation increases as the temperature decreases, in clear contrast to the Johnson-noise trend shown in Fig.~\ref{fig:HubbardBenchmark}\textbf{c}. This opposite trend already indicates that the dominant fluctuations at low temperatures are non-thermal. The QFI sum rule in Eq.~\eqref{eq:qfiNoiseSpectrum} makes this separation explicit by decomposing the total fluctuation into quantum and thermal parts. As indicated by the blue line in Fig.~\ref{fig:HubbardBenchmark}\textbf{d}, the QFI-extracted quantum fluctuations of the current rise rapidly upon cooling. In contrast, the difference between the total spectral integral and the QFI gives the thermal contribution, which decreases upon cooling and follows the same trend as the Johnson noise.

Quantum fluctuations from the noise spectrum not only qualitatively track the onset of quantum effects but can quantitatively witness entanglement through Eq.~\eqref{eq:QFIWitnessInequality}. At relatively high temperatures ($T>0.25t_h$), the quantum fluctuation of current remains below $\mu(1)$, the upper bound for separable states. In this regime, the measured current noise is not sufficient to certify entanglement. As the temperature is lowered, however, the strong increase of the QFI reflects the buildup of quantum correlations. For $T<0.25t_h$, the current-noise QFI crosses $\mu(1)$, which means that the observed current noise is too large to be generated by any separable state and therefore certifies the presence of at least bipartite entanglement. 

A spectrum-based entanglement witness requires quantitative accuracy in the spectral intensity, rather than agreement only in the qualitative line shape. This requirement raises a familiar concern for imaginary-time methods: analytic continuation generally does not yield a unique real-frequency spectrum\,\cite{JarrellGubernatis1996, GunnarssonEtAl2010}. This ambiguity, however, does not affect the entanglement-witness strategy developed here. The sum rule in Eq.~\eqref{eq:qfiNoiseSpectrum} can be reformulated as a fermionic-Matsubara representation in imaginary frequency, which imposes an exact constraint on the QFI (see \SISec\ IV). Therefore, $F_Q$ evaluated by DQMC is quantitatively reliable, insensitive to numerical uncertainties of analytical continuation.

\begin{figure*}[!t]
    \hspace*{-3mm}  \centering
    \includegraphics[width=183mm]{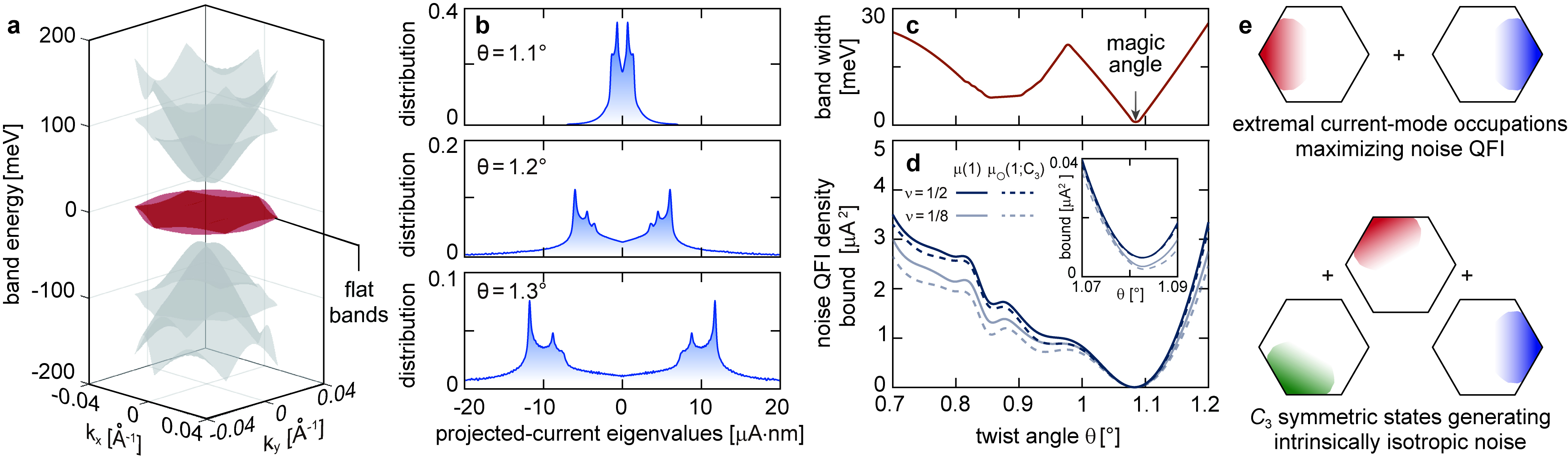}\vspace{-3mm}
    \caption{\textbf{Projected active space and isotropic noise in twisted bilayer graphene (TBG).} \textbf{a}, Low-energy band structure of TBG (twist angle $\theta = 1.2^\circ$), where the two red-highlighted bands identify the flat-band manifold effectively separated from remote bands. \textbf{b}, Distributions of projected-current eigenvalues at $\theta = 1.1^\circ$, $1.2^\circ$, and $1.3^\circ$. \textbf{c}, Twist-angle dependence of bandwidth. \textbf{d}, Twist-angle dependence of the bounds for noise QFI (solid) and $C_3$-symmetric isotropic noise (dashed) at $\nu=1/8$ and $1/2$ per spin and valley. The inset shows the region near the magic angle. \textbf{e}, Schematic illustration of the extremal current-mode occupations (upper) that maximize the noise QFI and define the supremum in Eq.~\eqref{eq:QFIBoundRes}, contrasted with the symmetry-constrained states (lower) that determine the suprema of the isotropic-noise bound in Eq.~\eqref{eq:isotropicQFIDef}.
    }
    \label{fig:TBGBound}
\end{figure*}

Because entanglement is most clearly witnessed at low temperatures, we next turn to the zero-temperature limit and analyze it using large-scale DMRG (see \textbf{Methods}). As shown in Fig.~\ref{fig:HubbardBenchmark}\textbf{e}, the noise QFI certifies entanglement only for $U>6t_h$, where the system is already deep in the Mott insulating phase. Below this threshold, the current fluctuations are not sufficiently strong to witness entanglement. As a benchmark, we also calculate the maximal eigenvalue $\lambda_{\max}$ of 2CRDM\,\cite{juhasz2006cumulant,schouten2022large}, which serves as a basis-independent entanglement metric for fermionic systems (see \textbf{Methods})\,\cite{liu2025entanglement}. As shown in Fig.~\ref{fig:HubbardBenchmark}\textbf{f}, the 2CRDM follows the same trend as the noise QFI. In the strong-coupling regime, it identifies at least 5-partite entanglement, while for the regime $U<6t_h$, where the noise QFI no longer witnesses entanglement, the 2CRDM still characterizes entangled states\,\cite{entanglementCondition}. Thus, the noise QFI is less sensitive than the 2CRDM witness. Its major advantage, however, lies in its experimental accessibility. Unlike the 2CRDM, which involves four independent indices and requires nonlinear spectroscopy\,\cite{liu2025entanglement}, current-noise QFI can be measured through a much simpler experimental approach, making it particularly promising for quantum materials under cryogenic conditions. 

\subsection{Projected and isotropic noise bounds}\label{sec:symmetry_constrained_bound}

The witness inequality and its sensitivity to entanglement depend on both the observable and the constrained state manifold, explaining the difference between the noise-QFI and 2CRDM witnesses. A natural route to tightening a witness inequality like Eq.~\eqref{eq:QFIWitnessInequality} is therefore to project the entanglement witness onto an active subspace, whenever the system exhibits a well-defined separation of energy scales. Since remote bands are nearly full or empty, including remote-band excitations in the integration of Eq.~\eqref{eq:qfiNoiseSpectrum} drives the bound in Eq.~\eqref{eq:QFIBoundRes} to an excessively large value, making the witness inequality unnecessarily loose. To avoid this, we reformulate the witness in the Fock space of an active-band manifold using an effective low-energy current operator. This projection is justified when the active bands are separated from the remote bands by a gap that is sufficiently larger than the active-band width, interaction strength, and temperature. Corrections from virtual remote-band processes can then be incorporated systematically through a Schrieffer-Wolff downfolding procedure (see \SISec\ V for details).  

Here, we illustrate the projected entanglement bounds using twisted bilayer graphene (TBG) as an example. Since the experimentally relevant correlated phases appear when the Fermi level lies within the two flat bands of each valley, we project the current operator onto this active two-band manifold (see Fig.~\ref{fig:TBGBound}\textbf{a}), as shown in Eq.~\eqref{eq:projectedCurrentEff} of \textbf{Methods}. The basis-independent $\mu(k)$ is then obtained by substituting the projected eigenvalues into Eq.~\eqref{eq:QFIBoundRes}. The resulting spectral radius of this projected current is set by the flat-band bandwidth rather than by the full bandwidth of the system. As shown in Fig.~\ref{fig:TBGBound}\textbf{b}, these eigenvalues become concentrated near zero as the twist angle approaches the magic angle, $\theta=1.08^\circ$, where the flat bands are narrowest.

As a result, the upper bound for separable states becomes especially small near the magic angle (see Figs.~\ref{fig:TBGBound}\textbf{c} and \textbf{d}). Notably, the maximal noise fluctuation is determined by the extremal current-mode occupations in Eq.~\eqref{eq:QFIBoundRes} (see Fig.~\ref{fig:TBGBound}\textbf{e}). These extremal configurations remain nearly unchanged over a broad filling range. Thus, the resulting upper bounds on the noise QFI depend only weakly on filling compared with their dependence on twist angle (see Fig.~\ref{fig:TBGBound}\textbf{d}).

\begin{figure*}[!t]
    \hspace*{-3mm}  \centering
    \includegraphics[width=183mm]{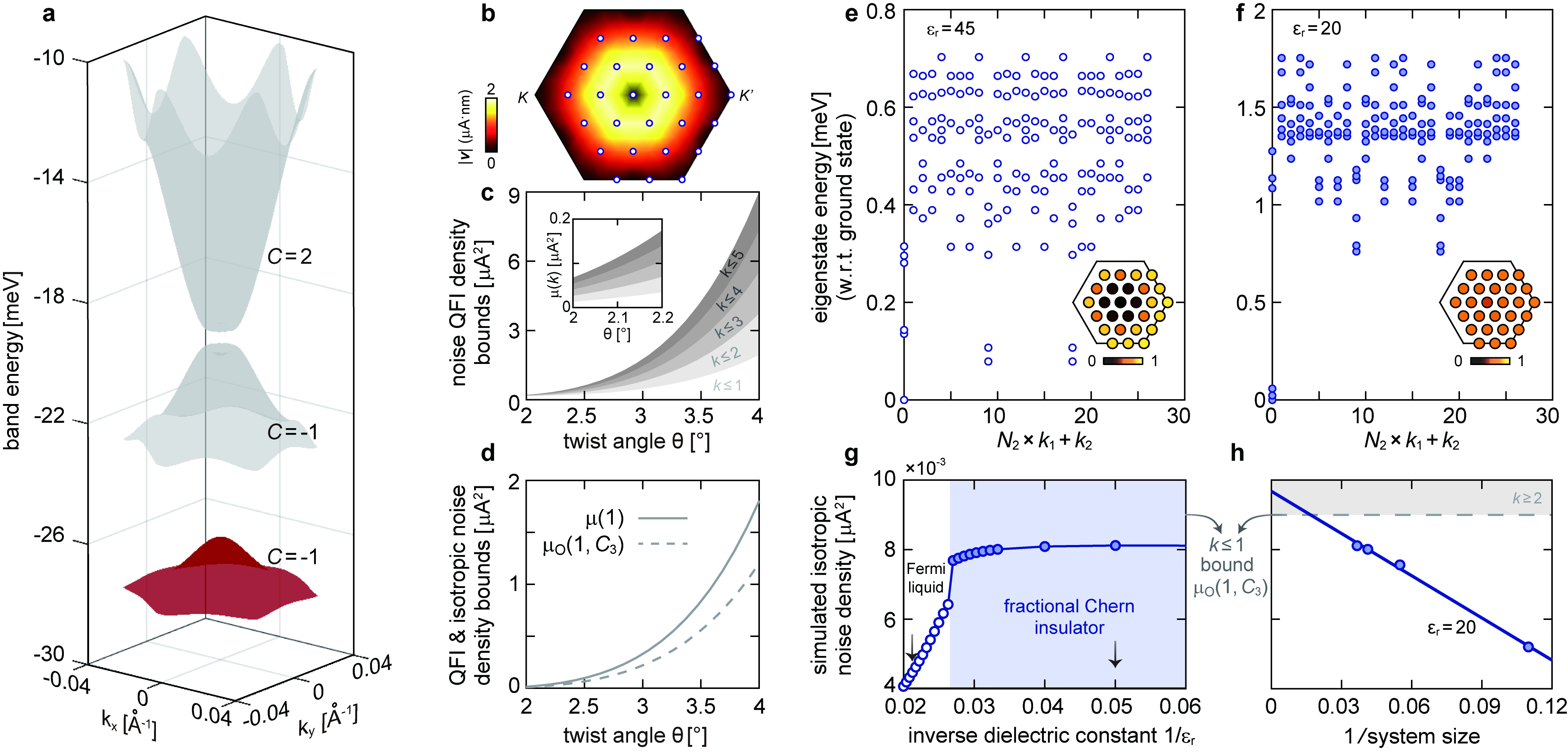}\vspace{-3mm}
   \caption{\textbf{Entanglement witness in twisted bilayer MoTe$_2$ (tMoTe$_2$) and fractional Chern insulator (FCI).}
   \textbf{a}, Low-energy band structure of tMoTe$_2$, with the projected Chern band highlighted. \textbf{b}, Projected current magnitude over the mini-Brillouin zone; circles denote the 27 momenta included in ED simulations. \textbf{c}, Twist-angle dependence of the upper bounds $\mu(k)$ for the noise QFI density at $\nu = 2/3$ hole filling per moir\'e unit cell, with the inset highlighting the region near the magic angle. \textbf{d}, Comparison between the $k=1$ noise-QFI bound (solid line, same as in \textbf{c}) and the isotropic-noise bound (dashed line). \textbf{e},\textbf{f}, Momentum-resolved eigenspectra, labeled by the linearized momentum index $N_2\times k_1 + k_2$, for the Fermi-liquid phase at $\varepsilon_r=45$ (\textbf{e}) and the FCI phase at $\varepsilon_r=20$ (\textbf{f}). Insets show the corresponding ground-state occupation $\langle n_{\mathbf{k}}\rangle$ across the Brillouin zone. \textbf{g}, Isotropic noise $G_\bigcirc$ for the 27-site cluster, calculated by ED as a function of dielectric constant $\varepsilon_r$, which controls the effective band-projected Coulomb interaction. The parameters corresponding to \textbf{e} and \textbf{f} are indicated. Results are shown for twist angle $\theta = 2^{\circ}$ and $\nu = 2/3$. The shaded region denotes the FCI phase. \textbf{h}, Finite-size scaling of $G_\bigcirc$ in the FCI phase at $\varepsilon_r=20$, for $N=9$, 18, 24, and 27. The line denotes a linear fit  over $1/N$. The arrows in \textbf{g} and \textbf{h} mark the upper bound for separable ($k=1$) states.}
    \label{fig:mote2FCI}
\end{figure*}


Although mathematically allowed, these atypical states with extremal current-mode occupations raise the supremum to a high value and thereby make Eq.~\eqref{eq:QFIBoundRes} less efficient. Therefore, a second route to tightening the witness inequality is to restrict the admissible wavefunctions by enforcing the system's physical symmetries, excluding highly polarized states that violate them (see Fig.~\ref{fig:TBGBound}\textbf{e}). For crystalline materials, the wavefunction manifold can be constrained by the appropriate rotational point-group symmetry. To incorporate this constraint and quantify the intrinsically isotropic noise, we average the QFIs along all angular directions,
\begin{equation}\label{eq:isotropicQFIDef}
    G_{\bigcirc}(T) = \frac{1}{\pi}\int_0^{\pi} F_Q[\hat{J}_\theta](T) d\theta\,,
\end{equation}
with $\theta$ denoting the measurement direction relative to the $x$ axis. This isotropic observable is accessible in angle-resolved transport measurements or polarization-dependent optical measurements on nanoscale devices.

The occupation construction underlying Eq.~\eqref{eq:QFIBoundRes} does not apply directly to $G_{\bigcirc}(T)$, because the extrema of the individual QFI components cannot be achieved simultaneously by the same state. Evaluating the supremum of the isotropic noise density, denoted by $\mu_{\bigcirc}(k;C_n)$, therefore requires optimization over admissible states transforming under any irreducible representation of $C_n$. The derivation and optimization procedure are given in \textbf{Methods} and \SISec\ VI. For TBG specifically, Fig.~\ref{fig:TBGBound}\textbf{d} shows that the symmetry-constrained supremum always lies below the corresponding (unconstrained) noise QFI bound. The resulting inequality, $G_{\bigcirc}(T)\leq \mu_{\bigcirc}(k;C_n)\cdot\mathcal{A}$, is therefore tighter than Eq.~\eqref{eq:QFIWitnessInequality}, giving a more efficient entanglement witness. A measured isotropic noise density exceeding $\mu_{\bigcirc}(k;C_n)$ for a $C_n$-symmetric system certifies a $(k+1)$-partite entangled state. In other words, the witness inequality is tightened because the same admissible state must supply all angular components and the independently largest fluctuations along different directions generally cannot be combined into one state. 

\subsection{Witnessing entanglement in Twisted {$\textrm{MoTe}_2$}}\label{sec:graphene_flatband_bounds}

We next apply the projected and isotropic noise bounds to the twisted bilayer MoTe$_2$ (tMoTe$_2$), which is a clean platform for realizing fractional Chern insulator (FCI) states at zero magnetic field\,\cite{devakul2021magic,li2021spontaneous, cai2023signatures, zeng2023thermodynamic,park2023observation}. Although fractionalized excitations are widely expected to indicate topological entanglement\,\cite{li2008entanglement}, direct certification of such entanglement in materials remains absent, which motivates the noise-based witness proposed here. To model tMoTe$_2$, we use the continuum Hamiltonian described in \textbf{Methods}. At twist angle $\theta=2^\circ$, this model hosts a nearly flat topological moir\'e band separated from the remote bands by a gap $\sim3\,{\rm meV}$. We project the Coulomb interaction onto this active band (see Fig.~\ref{fig:mote2FCI}\textbf{a})\,\cite{reddy2023fractional, wang2024fractional, sharma2024topological}, which preserves the $C_3$ rotational symmetry.

Within this projected-band framework, we evaluate the distribution of the projected current operators across the Brillouin zone (see Fig.~\ref{fig:mote2FCI}\textbf{b}). This distribution determines the noise-QFI bounds $\mu(k)$ through Eq.~\eqref{eq:QFIWitnessInequality} and \eqref{eq:QFIBoundRes}. We focus on hole filling $\nu=2/3$ per moir\'e unit cell, where the FCI phase is expected to be robust. The resulting bounds are shown in Fig.~\ref{fig:mote2FCI}\textbf{c}. As the twist angle increases from $2^\circ$ to $4^\circ$, $\mu(k)$ increases monotonically. This behavior follows from the increase in bandwidth, which enlarges the allowed range of current fluctuations. For comparison, we also evaluate the isotropic bounds $\mu_{\bigcirc}(k;C_n)$, now incorporating the $C_3$ rotational symmetry of tMoTe$_2$ (see Fig.~\ref{fig:mote2FCI}\textbf{d}). Once this symmetry is enforced, the upper bound is reduced by about 30\%. This substantial tightening of the bound lowers the experimental threshold for entanglement certification and shows that symmetry information can significantly enhance the sensitivity of noise-based entanglement detection.

With these bounds established, we calculate the ground state of the interacting Hamiltonian for tMoTe$_2$ using exact diagonalization (ED). To preserve the $C_3$ symmetry, we adopt a 27-site cluster containing the high-symmetry momenta shown in Fig.~\ref{fig:mote2FCI}\textbf{b}. The Coulomb interaction strength is tuned by the dielectric constant $\varepsilon_r$. In the weak-interaction limit, the many-body eigenspectrum remains gapless, and the Bloch state occupation $n_{\mathbf{k}}=\langle \gamma_{\mathbf{k}}^\dagger \gamma_{\mathbf{k}}\rangle$ exhibits a clear Fermi surface (see Fig.~\ref{fig:mote2FCI}\textbf{e}). These features are characteristic of a Fermi-liquid phase. As the interaction increases, the system enters the FCI regime at $\varepsilon_r \approx 37$ for this cluster size, where the ground state develops a threefold approximate degeneracy, well separated from excited states (see Fig.~\ref{fig:mote2FCI}\textbf{f}). At the same time, $n_{\mathbf{k}}$ becomes nearly uniform across the Brillouin zone. 

Importantly, the isotropic noise $G_{\bigcirc}$ tracks this transition closely and provides a direct indicator of the underlying phases. In the weak-interaction regime, the current fluctuations remain small, as shown in Fig.~\ref{fig:mote2FCI}\textbf{g}. Quantum fluctuations build up near the transition and increase sharply once the system enters the FCI phase. This enhancement is not confined to the critical region near $\varepsilon_r\approx 37$, as is often the case for fluctuation peaks near magnetic criticality\,\cite{mazza2026quantum}. Instead, $G_{\bigcirc}$ stays large throughout the entire FCI regime. This sustained enhancement indicates that the isotropic noise is not merely detecting proximity to a quantum criticality but is instead tracking the intrinsic correlations of the FCI state itself.

Because our calculation is performed on a finite cluster, it captures current fluctuations only at sampled high-symmetry momenta. Consequently, although the enhanced isotropic noise closely approaches the bound $\mu_{\bigcirc}(1;C_3)\approx 0.009 \,\mu \text{A}^2$, it remains below this bound for the 27-site cluster. To determine the expected behavior in the thermodynamic limit, we therefore perform finite-size scaling of the simulated isotropic noise $G_{\bigcirc}$ across different system sizes using a linear fit in $1/N$. As shown in Fig.~\ref{fig:mote2FCI}\textbf{h}, the extrapolated isotropic noise density exceeds the upper bound for separable states in the thermodynamic limit ($N\rightarrow\infty$). Despite the uncertainty associated with the finite-size extrapolation, the extrapolated intercept remains above $\mu_{\bigcirc}(1;C_3)$ at the 95\% confidence interval of (9.1 -- 10.1)$\times 10^{-3}\,\mu\text{A}^2$. This result indicates that transport-noise or optical measurements in tMoTe$_2$ should enable direct certification of entanglement in experiments.

\section{Discussion}

Our results establish quantum fluctuations encoded in the noise spectrum as a practical witness of entanglement in quantum materials, particularly for non-magnetic materials and systems that are difficult to access using scattering experiments because of small sizes, weak cross-sections, or low-temperature requirements. The corresponding witness inequalities can be substantially tightened by incorporating rotational symmetry. The isotropic noise can be reconstructed from angle-resolved transport or polarization-dependent optical measurements, with irreducible representations of the symmetry reducing the required angular sampling of $F_Q[\hat{J}_\theta](T)$ to a few directions. Additional symmetries, including translational and valley symmetry, offer further systematic routes for improving the sensitivity of the witness.

Experimental measurements inevitably capture current fluctuations only within a finite frequency window. This limitation further highlights the suitability of the transport-based protocol for flat-band materials: their small intrinsic bandwidth provides a naturally low energy cutoff, with an isolated active-band manifold, and the spectral integral can be restricted to low-energy excitations below the onset of remote-band transitions (see \SISec\ V). This selection excludes unrelated high-frequency contributions and allows the use of tighter entanglement-witness bounds (see Fig.~\ref{fig:TBGBound}), enhancing sensitivity to weak fluctuation signals even with a partial sum rule. Information beyond the transport bandwidth (usually up to GHz range) can be supplemented by terahertz spectroscopy and optical reflectivity, which constrain high-frequency current response through the fluctuation--dissipation theorem. In the low-temperature limit, the kernel in Eq.~\eqref{eq:qfiNoiseSpectrum} approaches unity and fluctuations are predominantly quantum. The witness then no longer requires a complete Fourier reconstruction of Eq.~\eqref{eq:symmetrizedNoiseSpec}; instead, the sum rule reduces to the equal-time current fluctuation $4\langle \delta \hat{J}_\alpha^2\rangle(T)$ (see Fig.~\ref{fig:HubbardBenchmark}\textbf{d}), which can be measured as a correlation function without resolving its spectral distribution.

This work focuses on equilibrium current-noise fluctuations. Extending the framework to finite bias voltages is a promising direction, particularly because shot-noise measurements have already revealed signatures of pairing, fractionalization, and non-Fermi-liquid physics\,\cite{saminadayar1997observation, zhou2019electron, chen2023shot, niu2024shot}. Beyond the FCI state in tMoTe$_2$ studied here, candidate platforms include strange metals in moir\'e systems, correlated quantum-dot systems, and the FCI and chiral superconducting phases of rhombohedral graphene.


\section{Methods}

\subsection{Hubbard model}\label{sec:methodsHubbard}

The Hubbard-model simulations presented in Fig.~\ref{fig:HubbardBenchmark} use the single-band Hamiltonian
\begin{equation}\label{eq:hubbard}
    \mathcal{H} = -t_h\sum_{\langle ij\rangle,\sigma} c^\dagger_{i\sigma}c_{j\sigma} +U\sum_i\left(n_{i\uparrow}-\frac{1}{2}\right)\left(n_{i\downarrow}-\frac{1}{2}\right)\,,
\end{equation}
where $t_h$ is the nearest-neighbor hopping integral and $U$ is the on-site interaction. The interaction with external field is incorporated through the Peierls substitution $t_h\rightarrow t_h\exp[-ie\mathbf A\cdot(\mathbf r_i-\mathbf r_j)/\hbar]$.  The paramagnetic current in the $x$ direction is
\begin{equation}
    \hat J_x=\ii \frac{ e a_0 t_h }{\hbar}\sum_{i,\sigma}\left(c^\dagger_{i\sigma}c_{i+1_x,\sigma}-c^\dagger_{i+1_x,\sigma}c_{i\sigma}\right)\,.
\end{equation}
The $k$-producible thresholds in the Hubbard figures are obtained by applying the sorted-spectrum construction in Eq.~\eqref{eq:QFIBoundRes} to this spin-resolved current spectrum by diagonalizing $\hat J_x$. The lattice constant is algebraically denoted as $a_0$, including a single unit-cell spread along the $y$ direction, which is used in the evaluation of QFI density in Eq.~\eqref{eq:QFIBoundRes}.

\subsection{DQMC simulations}

The finite-temperature spectra and QFIs are simulated using DQMC\,\cite{blankenbecler1981monte,assaad2008world}. A discrete Hubbard--Stratonovich transformation first maps the interacting problem onto noninteracting fermions coupled to fluctuating auxiliary fields, whose configurations are sampled by Monte Carlo. On a finite lattice, DQMC is numerically exact up to controllable imaginary-time discretization and statistical errors. The calculations shown in Figs.~\ref{fig:HubbardBenchmark}\textbf{a}--\textbf{d} are performed for the half-filled Hubbard model with interaction strength $U=8t_h$ on a 40-site chain with periodic boundary conditions. At half filling, particle--hole symmetry renders the simulations free of the fermion sign problem. We use an imaginary-time step $\Delta\tau=0.05\,t_h^{-1}$ and perform 1,120 independent Monte Carlo runs. In each run, the first 500 sweeps are discarded for thermalization, followed by 1,000 measurement sweeps, which are grouped into bins of 200 sweeps for statistical analysis.

The simulations directly measure the imaginary-time current correlation function
\begin{equation}\label{eq:method_Lambda_tau}
    \Lambda(\tau) = \left\langle \hat J_x(\tau)\hat J_x(0)\right\rangle, \qquad0\leq\tau\leq\beta.
\end{equation}
To obtain the symmetrized current-noise spectra $S_x(\omega,T)$ shown in Figs.~\ref{fig:HubbardBenchmark}\textbf{a} and \textbf{b}, we analytically continue $\Lambda(\tau)$ using maximum entropy\,\cite{JarrellGubernatis1996}. The QFI values in Fig.~\ref{fig:HubbardBenchmark}\textbf{d} are then obtained from the weighted spectral integral in Eq.~\eqref{eq:qfiNoiseSpectrum}. Although analytic continuation is ill-conditioned for noisy imaginary-time data, the integrated QFI is substantially more robust than the detailed real-frequency line shape. As shown in \SISec\ IV, causality allows the QFI integral to be rewritten exactly as a fermionic-Matsubara sum on the imaginary-frequency axis, explaining this robustness.

\subsection{DMRG simulations}
Zero-temperature QFI and 2CRDM results for the one-dimensional Hubbard model are simulated using DMRG. The calculations are performed on open chains of length $L=160$ at half-filling, with both the total particle number and total magnetization fixed, the latter to zero. We impose a singular-value cutoff of $10^{-7}$ during truncation and retain bond dimensions up to $M=675$. Finite-system sweeps are continued until the ground-state energy and all observables relevant to the entanglement analysis are converged.

The converged matrix product state is then used to evaluate two entanglement witnesses. The noise QFI is computed directly from the static current correlation function $F_Q(\hat J_{\alpha})=4\langle\delta\hat J_{\alpha}^2\rangle$ at zero temperature. The 2CRDM is constructed using equal-time four-point and two-point correlation functions, following Eq.~\eqref{eq:2crdm}. To minimize boundary effects, these correlations are measured only within the central $36$ sites of the chain. This restriction also ensures that the four-index 2CRDM cumulant tensor remains numerically tractable, because both its storage and measurement costs scale with the fourth power of the number of included sites.

\subsection{2CRDM and entanglement witness}
The 2CRDM is defined as the connected part of the two-particle reduced density matrix\,\cite{juhasz2006cumulant,schouten2022large},
\begin{equation}\label{eq:2crdm}
\Gamma_{\mu\nu,\mu'\nu'}=\langle c_\mu^\dagger c_\nu^\dagger c_{\nu'} c_{\mu'}\rangle -\langle c_\mu^\dagger c_{\mu'}\rangle \langle c_\nu^\dagger c_{\nu'}\rangle + \langle c_\mu^\dagger c_{\nu'}\rangle  \langle c_\nu^\dagger c_{\mu'}\rangle \,,
\end{equation}
where the single-particle basis labels $\mu$ and $\nu$ include both site and spin indices. We flatten this tensor into the matrix $\Gamma_{(\mu\mu'),(\nu\nu')}$ and use its largest eigenvalue, $\lambda_{\rm max}$, as a basis-independent entanglement witness. As proven in Ref.~\onlinecite{liu2025entanglement}, electronic states with entanglement depth no greater than $k$ satisfy the upper bound $\left( k - {1}/{2} \right)/2$, for any integer $k>1$. Accordingly, the results in Fig.~\ref{fig:HubbardBenchmark}\textbf{f} certify at least $(k+1)$-partite entanglement whenever $\lambda_{\rm max}$ exceeds the bound associated with entanglement depth $k$.

\subsection{Projected model for twisted bilayer graphene}\label{supp:graphene_flatband_models}
The single-particle electronic structure of twisted bilayer graphene used in Fig.~\ref{fig:TBGBound} is constructed using the Bistritzer--MacDonald model\,\cite{bistritzer2011moire}. Spin is treated as a passive degeneracy, and the Hamiltonian is considered independently in each valley sector $\xi=\pm1$, with intervalley scattering neglected. For each valley, the continuum Hamiltonian matrix is
\begin{equation}\label{eq:BMmodelTBG}
    H_{\xi}(\mathbf{k},\mathbf{r}) = \begin{pmatrix} h_{\xi,+\theta/2}(\mathbf{k}) & T_{\xi}(\mathbf{r}) \\ T_{\xi}^{\dagger}(\mathbf{r}) & h_{\xi,-\theta/2}(\mathbf{k}) \end{pmatrix}\,,
\end{equation}
where the intralayer terms
\begin{equation}
    h_{\xi,\pm\theta/2}(\mathbf{k}) = -\hbar v_F \left[ R(\pm\theta/2) \left( \mathbf{k}-\mathbf K_{\xi}^{(\pm)} \right) \right] \cdot (\xi\sigma_x,\sigma_y)
\end{equation}
describe the Dirac Hamiltonians of the top and bottom graphene layers rotated by $\pm\theta/2$, respectively. Here, $R(\varphi)$ is the in-plane rotation matrix, and $\mathbf K_{\xi}^{(\pm)}$ denote the valley centers of the corresponding rotated layers.

The interlayer moir\'e tunneling is given by
\begin{equation}
    T_\xi(\mathbf r) = \sum_{j=1}^{3} T_j^{(\xi)} e^{\ii\xi\mathbf q_j\cdot\mathbf r},
\end{equation}
where the three vectors $\mathbf q_j$ connect the Dirac points of the two graphene layers. The corresponding tunneling matrices are
\begin{equation}
    T_j^{(\xi)} = w_0\sigma_0 + w_1 \left[ \cos(\xi\varphi_j)\sigma_x + \sin(\xi\varphi_j)\sigma_y \right],
\end{equation}
with $\varphi_1=0$, $\varphi_2=2\pi/3$, and $\varphi_3=-2\pi/3$. The amplitudes $w_0$ and $w_1$ describe tunneling between locally AA- and AB/BA-stacked regions, respectively. We take the carbon--carbon bond length to be $d=1.42~\mathrm{\AA}$, the monolayer Dirac coefficient to be $\hbar v_F=5.25\,\mathrm{eV\cdot\AA}$, and the interlayer tunneling amplitudes to be $w_0=79.7~\mathrm{meV}$ and $w_1=97.5~\mathrm{meV}$. The condition $w_0<w_1$ accounts for the leading effects of lattice relaxation and corrugation, which reduce AA tunneling relative to AB/BA\,\cite{koshino2018maximally}.

\subsection{Model and simulations for tMoTe$_2$}\label{sec:mote2}
The valence-band edges of monolayer MoTe$_2$ lie at the $\textbf{K}$ and $\textbf{K}'$ valleys, where strong spin-orbit coupling locks the spin and valley degrees of freedom\,\cite{xiao2012coupled}. Consequently, for a fixed valley in an AA-stacked twisted bilayer, the two layers carry the same spin species and can hybridize through spin-conserving interlayer tunneling. A small twist angle produces moir\'e-periodic intralayer potentials and spatially modulated interlayer tunneling, leading to the following spin-up single-particle Hamiltonian matrix\,\cite{wu2019topological}:
\begin{equation}\label{eq:BMmodelMoTe2}
    h(\mathbf{k}) = \begin{pmatrix} h_t+V_t(\mathbf{r})  & T(\mathbf{r}) \\     T^\dagger(\mathbf{r}) &  h_b+ V_b(\mathbf{r})    \end{pmatrix}\,.
\end{equation}
Here, $h_{t/b} = {\hbar^2 (\mathbf{k}-\mathbf{k}_{t/b})^2}/{2m^*}$ describe the kinetic Hamiltonians of the top ($t$) and bottom ($b$) layers, while $V_{t/b}(\mathbf{r}) = -2V_0\sum_{i=1,3,5} \cos(\mathbf{g_i}\cdot{\mathbf{r}} + \phi_{t/b})$ are the corresponding intralayer moir\'e potentials. The reciprocal moir\'e lattice vectors are defined by $\mathbf{g}_i = \frac{4\pi}{\sqrt{3}a_M} \big( \cos({\pi(i-1)}/{3}), \sin({\pi(i-1)}/{3}) \big)$, where $a_M$ is the moir\'e periodicity. The momenta $\mathbf{k}_{t/b}$ denote the layer-shifted $K$ points located at the mini-Brillouin-zone corners. The $C_{3z}$-symmetric interlayer tunneling is $T(\mathbf r)=w\left(1+e^{i\mathbf g_2\cdot\mathbf r}+e^{i\mathbf g_3\cdot\mathbf r}\right)$. We take $m^*\simeq0.62m_e$, with $m_e$ the bare electron mass. The remaining continuum-model parameters are material-dependent and can be obtained from first-principles calculations. For the results presented in Fig.~\ref{fig:mote2FCI}, we adopt $V_0 = 11.2\text{ meV}$, $\phi_{t/b} = \mp 91^\circ$, and $w = -13.3\text{ meV}$ from Ref.~\onlinecite{reddy2023fractional}. 

To simulate the many-body ground state of tMoTe$_2$, we include the Coulomb interaction
\begin{equation}
    \mathcal{H}_{\mathrm{int}}= \frac{1}{2\mathcal{A}}\sum_{\substack{\tau\tau' ll'\\ \mathbf{k}\mathbf{k}'\mathbf{q}}} V(\mathbf{q})\, c^\dagger_{\tau l,\mathbf{k}+\mathbf{q}}\, c^\dagger_{\tau' l',\mathbf{k}'-\mathbf{q}}\, c_{\tau' l',\mathbf{k}'}\,
c_{\tau l,\mathbf{k}}\, ,
\label{eq:interaction}
\end{equation}
where $V(\mathbf{q})=2\pi e^2/(\varepsilon_r|\mathbf{q}|)$ is the unscreened Coulomb potential and $\mathcal{A}={\sqrt{3}a_0^2N}/{8\sin^2(\theta/2)}$ is the total area of the finite moir\'e cluster. Here, $N$ is the number of discrete momentum points and $a_0=3.52\,\text{\AA}$ is the lattice constant of monolayer MoTe$_2$. The operator $c^\dagger_{\tau l,\mathbf{k}}$ creates an electron with momentum $\mathbf{k}$ in valley/spin $\tau$ and layer $l$. 

In the many-body simulation, we first diagonalize the single-particle continuum Hamiltonian in a plane-wave basis
\begin{equation}\label{eq:singleParticleBasisMoTe2}
    h(\mathbf{k})\left|u_{n\mathbf{k}}\right\rangle = \varepsilon_{n\mathbf{k}} \left|u_{n\mathbf{k}}\right\rangle .
\end{equation}
The interacting Hamiltonian in Eq.~\eqref{eq:interaction} is projected onto the lowest topological moir\'e band, labeled by $n=0$. Within this projected subspace, the plane-wave and band operators are related through $c_{\tau l,\mathbf{k}}= u_{0\tau l}(\mathbf{k})\,\gamma_{\mathbf{k}}$, where $u_{0\tau l}(\mathbf{k})$ is the lowest-band Bloch eigenvector and $\gamma_{\mathbf{k}}$ annihilates an electron in the projected band. The resulting interacting Hamiltonian is solved by exact diagonalization using a parallel Arnoldi algorithm on clusters with $N=9$, 18, 24, and 27 sites, all containing high symmetry points. Since the ground state is fully polarized\,\cite{reddy2023fractional, sharma2024topological}, we retain only a single valley/spin component.

\subsection{Projection of the current operator onto active moir\'e bands}
\label{sec:projected_current}
The projected current operator is formulated within a unified framework for TBG and tMoTe$_2$. In the Bloch basis, projection onto the active-band manifold gives
\begin{equation}\label{eq:projectedCurrentMatrix}
    \left[J_\alpha^{\rm (proj)}(\mathbf{k})\right]_{mn} = \frac{e}{\hbar} \left\langle u_{m\mathbf{k}} \left| \frac{\partial h(\mathbf{k})}{\partial k_\alpha} \right| u_{n\mathbf{k}} \right\rangle .
\end{equation}
The structure of the resulting operator depends on the number of bands retained in the active manifold. For TBG, two active bands are included, so $J_\alpha^{\rm (proj)}(\mathbf{k})$ is a $2\times2$ Hermitian matrix. Diagonalizing it at each momentum defines the projected-current eigenbasis,
\begin{equation}\label{eq:projectedCurrentEff}
    \hat J_\alpha^{\rm (proj)} = \sum_{\mathbf{k}\sigma} \sum_{m=1}^{2} \widetilde v_{m\alpha}(\mathbf{k})\, d_{\mathbf{k}\sigma m}^{\dagger} d_{\mathbf{k}\sigma m}\, ,
\end{equation}
where $\widetilde v_{m\alpha}(\mathbf{k})$ denote the corresponding projected-current eigenvalues. Substituting them into Eq.~\eqref{eq:QFIBoundRes} yields the projected entanglement bounds. For tMoTe$_2$, the active manifold contains an isolated topological moir\'e band. The projected current matrix therefore reduces to the scalar current $\widetilde v_\alpha(\mathbf{k})={(e/\hbar)\partial\varepsilon(\mathbf{k})}/{\partial k_\alpha}$. These projected current modes are used to calculate the entanglement bounds shown in Figs.~\ref{fig:mote2FCI}\textbf{c} and \textbf{d}. In the FCI phase, the current-current correlation function is evaluated independently for each of the three nearly degenerate ground states and then averaged, although the individual values are already nearly identical.

\subsection{Isotropic noise optimization}

With rotational symmetry, obtaining the upper bound on the isotropic noise $G_{\bigcirc}(T)$ for a $k$-producible state becomes a numerical optimization problem (see \SISec\ VI). We illustrate this procedure for the separable case, $k=1$. Let $\gamma$ be the one-particle reduced density matrix (1RDM) $\gamma_{ij} = \mel{\Psi_{1\text{-prod}}}{c_j^\dagger c_i}{\Psi_{1\text{-prod}}}$. For a given state $|\Psi_{1\text{-prod}}\rangle$, the isotropic noise is
\begin{equation}\label{eq:isotropicNoiseRDMDecomp}
    G_{\bigcirc}[\gamma] = 2\mathrm{Tr} \left[ \gamma J_x(1-\gamma)J_x +\gamma J_y(1-\gamma)J_y \right].
\end{equation}
Under $C_n$ symmetry, the 1RDM decomposes into irreducible angular-momentum sectors, $\gamma=\bigoplus_{\ell=0}^{n-1}\gamma^{(\ell)}$. The supremum $\mu_{\bigcirc}(1;C_n)$, introduced to distinguish it from the general result in Eq.~\eqref{eq:QFIWitnessInequality}, is therefore defined by optimizing $G_{\bigcirc}[\gamma]$ over all admissible sector 1RDM ${\gamma^{(\ell)}}$.

The optimization is conveniently reformulated using the circular current operators $\hat{J}_\pm=\hat{J}_x\pm \ii \hat{J}_y$. We denote their blocks connecting sectors $\ell\pm1$ and $\ell$ by $\hat{J}_\pm^{(\ell\pm1,\ell)}$. In this representation, the isotropic noise takes the form
\begin{eqnarray}\label{eq:isotropicNoiseRDMAngularDecomp}
    G_{\bigcirc}[\gamma] &=&
    \sum_{\ell=0}^{n-1}
    \mathrm{Tr}\qty[\gamma^{(\ell)} \left(J_-^{(\ell,\ell+1)}J_+^{(\ell+1,\ell)} + J_+^{(\ell,\ell-1)}J_-^{(\ell-1,\ell)}    \right)]\nonumber\\
    &&-    2\sum_{\ell=0}^{n-1}  \mathrm{Tr} \left( J_-^{(\ell,\ell+1)}  \gamma^{(\ell+1)}    J_+^{(\ell+1,\ell)} \gamma^{(\ell)}  \right).
\end{eqnarray}
The derivation of Eq.~\eqref{eq:isotropicNoiseRDMAngularDecomp} and the numerical optimization procedure are presented in \SISec\ VI.

\section*{Data availability}
The data that support the findings of this study are available from the shared \href{https://drive.google.com/drive/folders/1Yu-_mSSb8YBltVris4zqIEISJS0rTe7z}{Google Drive folder}, which will be converted into a permanent DOI before publications.

\section*{Code availability}
The code supporting this study is available at \href{https://drive.google.com/drive/folders/1Yu-_mSSb8YBltVris4zqIEISJS0rTe7z}{Google Drive folder}, which will be converted into a permanent DOI before publication.

\bibliography{refList_formatted}

\begin{thebibliography}{63}%
\makeatletter
\providecommand \@ifxundefined [1]{%
 \@ifx{#1\undefined}
}%
\providecommand \@ifnum [1]{%
 \ifnum #1\expandafter \@firstoftwo
 \else \expandafter \@secondoftwo
 \fi
}%
\providecommand \@ifx [1]{%
 \ifx #1\expandafter \@firstoftwo
 \else \expandafter \@secondoftwo
 \fi
}%
\providecommand \natexlab [1]{#1}%
\providecommand \enquote  [1]{``#1''}%
\providecommand \bibnamefont  [1]{#1}%
\providecommand \bibfnamefont [1]{#1}%
\providecommand \citenamefont [1]{#1}%
\providecommand \href@noop [0]{\@secondoftwo}%
\providecommand \href [0]{\begingroup \@sanitize@url \@href}%
\providecommand \@href[1]{\@@startlink{#1}\@@href}%
\providecommand \@@href[1]{\endgroup#1\@@endlink}%
\providecommand \@sanitize@url [0]{\catcode `\\12\catcode `\$12\catcode `\&12\catcode `\#12\catcode `\^12\catcode `\_12\catcode `\%12\relax}%
\providecommand \@@startlink[1]{}%
\providecommand \@@endlink[0]{}%
\providecommand \url  [0]{\begingroup\@sanitize@url \@url }%
\providecommand \@url [1]{\endgroup\@href {#1}{\urlprefix }}%
\providecommand \urlprefix  [0]{URL }%
\providecommand \Eprint [0]{\href }%
\providecommand \doibase [0]{https://doi.org/}%
\providecommand \selectlanguage [0]{\@gobble}%
\providecommand \bibinfo  [0]{\@secondoftwo}%
\providecommand \bibfield  [0]{\@secondoftwo}%
\providecommand \translation [1]{[#1]}%
\providecommand \BibitemOpen [0]{}%
\providecommand \bibitemStop [0]{}%
\providecommand \bibitemNoStop [0]{.\EOS\space}%
\providecommand \EOS [0]{\spacefactor3000\relax}%
\providecommand \BibitemShut  [1]{\csname bibitem#1\endcsname}%
\let\auto@bib@innerbib\@empty
\bibitem [{\citenamefont {Keimer}\ and\ \citenamefont {Moore}(2017)}]{keimer2017physics}%
  \BibitemOpen
  \bibfield  {author} {\bibinfo {author} {\bibfnamefont {B.}~\bibnamefont {Keimer}}\ and\ \bibinfo {author} {\bibfnamefont {J.~E.}\ \bibnamefont {Moore}},\ }\bibfield  {title} {\bibinfo {title} {{\textit{The Physics of Quantum Materials}}},\ }\href@noop {} {\bibfield  {journal} {\bibinfo  {journal} {Nat. Phys.}\ }\textbf {\bibinfo {volume} {13}},\ \bibinfo {pages} {1045} (\bibinfo {year} {2017})}\BibitemShut {NoStop}%
\bibitem [{\citenamefont {Basov}\ \emph {et~al.}(2017)\citenamefont {Basov}, \citenamefont {Averitt},\ and\ \citenamefont {Hsieh}}]{basov2017towards}%
  \BibitemOpen
  \bibfield  {author} {\bibinfo {author} {\bibfnamefont {D.}~\bibnamefont {Basov}}, \bibinfo {author} {\bibfnamefont {R.}~\bibnamefont {Averitt}},\ and\ \bibinfo {author} {\bibfnamefont {D.}~\bibnamefont {Hsieh}},\ }\bibfield  {title} {\bibinfo {title} {{\textit{Towards Properties on Demand in Quantum Materials}}},\ }\href@noop {} {\bibfield  {journal} {\bibinfo  {journal} {Nat. Mater.}\ }\textbf {\bibinfo {volume} {16}},\ \bibinfo {pages} {1077} (\bibinfo {year} {2017})}\BibitemShut {NoStop}%
\bibitem [{\citenamefont {Balents}(2010)}]{balents2010spin}%
  \BibitemOpen
  \bibfield  {author} {\bibinfo {author} {\bibfnamefont {L.}~\bibnamefont {Balents}},\ }\bibfield  {title} {\bibinfo {title} {{\textit{Spin Liquids in Frustrated Magnets}}},\ }\href@noop {} {\bibfield  {journal} {\bibinfo  {journal} {Nature}\ }\textbf {\bibinfo {volume} {464}},\ \bibinfo {pages} {199} (\bibinfo {year} {2010})}\BibitemShut {NoStop}%
\bibitem [{\citenamefont {Bernevig}\ \emph {et~al.}(2025)\citenamefont {Bernevig}, \citenamefont {Fu}, \citenamefont {Ju}, \citenamefont {MacDonald}, \citenamefont {Mak},\ and\ \citenamefont {Shan}}]{bernevig2025fractional}%
  \BibitemOpen
  \bibfield  {author} {\bibinfo {author} {\bibfnamefont {B.}~\bibnamefont {Bernevig}}, \bibinfo {author} {\bibfnamefont {L.}~\bibnamefont {Fu}}, \bibinfo {author} {\bibfnamefont {L.}~\bibnamefont {Ju}}, \bibinfo {author} {\bibfnamefont {A.}~\bibnamefont {MacDonald}}, \bibinfo {author} {\bibfnamefont {K.}~\bibnamefont {Mak}},\ and\ \bibinfo {author} {\bibfnamefont {J.}~\bibnamefont {Shan}},\ }\bibfield  {title} {\bibinfo {title} {{\textit{Fractional Quantization in Insulators from Hall to Chern}}},\ }\href@noop {} {\bibfield  {journal} {\bibinfo  {journal} {Nat. Phys.}\ }\textbf {\bibinfo {volume} {21}},\ \bibinfo {pages} {1702} (\bibinfo {year} {2025})}\BibitemShut {NoStop}%
\bibitem [{\citenamefont {Bergholtz}\ and\ \citenamefont {Liu}(2013)}]{bergholtz2013topological}%
  \BibitemOpen
  \bibfield  {author} {\bibinfo {author} {\bibfnamefont {E.~J.}\ \bibnamefont {Bergholtz}}\ and\ \bibinfo {author} {\bibfnamefont {Z.}~\bibnamefont {Liu}},\ }\bibfield  {title} {\bibinfo {title} {{\textit{Topological Flat Band Models and Fractional Chern Insulators}}},\ }\href@noop {} {\bibfield  {journal} {\bibinfo  {journal} {Int. J. Mod. Phys. B}\ }\textbf {\bibinfo {volume} {27}},\ \bibinfo {pages} {1330017} (\bibinfo {year} {2013})}\BibitemShut {NoStop}%
\bibitem [{\citenamefont {Greene}\ \emph {et~al.}(2020)\citenamefont {Greene}, \citenamefont {Mandal}, \citenamefont {Poniatowski},\ and\ \citenamefont {Sarkar}}]{greene2020strange}%
  \BibitemOpen
  \bibfield  {author} {\bibinfo {author} {\bibfnamefont {R.~L.}\ \bibnamefont {Greene}}, \bibinfo {author} {\bibfnamefont {P.~R.}\ \bibnamefont {Mandal}}, \bibinfo {author} {\bibfnamefont {N.~R.}\ \bibnamefont {Poniatowski}},\ and\ \bibinfo {author} {\bibfnamefont {T.}~\bibnamefont {Sarkar}},\ }\bibfield  {title} {\bibinfo {title} {{\textit{The Strange Metal State of the Electron-Doped Cuprates}}},\ }\href@noop {} {\bibfield  {journal} {\bibinfo  {journal} {Annu. Rev. Conden. Ma. P.}\ }\textbf {\bibinfo {volume} {11}},\ \bibinfo {pages} {213} (\bibinfo {year} {2020})}\BibitemShut {NoStop}%
\bibitem [{\citenamefont {Keimer}\ \emph {et~al.}(2015)\citenamefont {Keimer}, \citenamefont {Kivelson}, \citenamefont {Norman}, \citenamefont {Uchida},\ and\ \citenamefont {Zaanen}}]{keimer2015quantum}%
  \BibitemOpen
  \bibfield  {author} {\bibinfo {author} {\bibfnamefont {B.}~\bibnamefont {Keimer}}, \bibinfo {author} {\bibfnamefont {S.~A.}\ \bibnamefont {Kivelson}}, \bibinfo {author} {\bibfnamefont {M.~R.}\ \bibnamefont {Norman}}, \bibinfo {author} {\bibfnamefont {S.}~\bibnamefont {Uchida}},\ and\ \bibinfo {author} {\bibfnamefont {J.}~\bibnamefont {Zaanen}},\ }\bibfield  {title} {\bibinfo {title} {{\textit{From Quantum Matter to High-Temperature Superconductivity in Copper Oxides}}},\ }\href@noop {} {\bibfield  {journal} {\bibinfo  {journal} {Nature}\ }\textbf {\bibinfo {volume} {518}},\ \bibinfo {pages} {179} (\bibinfo {year} {2015})}\BibitemShut {NoStop}%
\bibitem [{\citenamefont {H{\"u}fner}\ \emph {et~al.}(2008)\citenamefont {H{\"u}fner}, \citenamefont {Hossain}, \citenamefont {Damascelli},\ and\ \citenamefont {Sawatzky}}]{hufner2008two}%
  \BibitemOpen
  \bibfield  {author} {\bibinfo {author} {\bibfnamefont {S.}~\bibnamefont {H{\"u}fner}}, \bibinfo {author} {\bibfnamefont {M.}~\bibnamefont {Hossain}}, \bibinfo {author} {\bibfnamefont {A.}~\bibnamefont {Damascelli}},\ and\ \bibinfo {author} {\bibfnamefont {G.}~\bibnamefont {Sawatzky}},\ }\bibfield  {title} {\bibinfo {title} {{\textit{Two Gaps Make a High-Temperature Superconductor}}},\ }\href@noop {} {\bibfield  {journal} {\bibinfo  {journal} {Rep. Prog. Phys.}\ }\textbf {\bibinfo {volume} {71}},\ \bibinfo {pages} {062501} (\bibinfo {year} {2008})}\BibitemShut {NoStop}%
\bibitem [{\citenamefont {Aspect}\ \emph {et~al.}(1982)\citenamefont {Aspect}, \citenamefont {Dalibard},\ and\ \citenamefont {Roger}}]{aspect1982experimental}%
  \BibitemOpen
  \bibfield  {author} {\bibinfo {author} {\bibfnamefont {A.}~\bibnamefont {Aspect}}, \bibinfo {author} {\bibfnamefont {J.}~\bibnamefont {Dalibard}},\ and\ \bibinfo {author} {\bibfnamefont {G.}~\bibnamefont {Roger}},\ }\bibfield  {title} {\bibinfo {title} {{\textit{Experimental Test of {B}Ell'S Inequalities Using Time-Varying Analyzers}}},\ }\href@noop {} {\bibfield  {journal} {\bibinfo  {journal} {Phys. Rev. Lett.}\ }\textbf {\bibinfo {volume} {49}},\ \bibinfo {pages} {1804} (\bibinfo {year} {1982})}\BibitemShut {NoStop}%
\bibitem [{\citenamefont {Est{\`e}ve}\ \emph {et~al.}(2008)\citenamefont {Est{\`e}ve}, \citenamefont {Gross}, \citenamefont {Weller}, \citenamefont {Giovanazzi},\ and\ \citenamefont {Oberthaler}}]{esteve2008squeezing}%
  \BibitemOpen
  \bibfield  {author} {\bibinfo {author} {\bibfnamefont {J.}~\bibnamefont {Est{\`e}ve}}, \bibinfo {author} {\bibfnamefont {C.}~\bibnamefont {Gross}}, \bibinfo {author} {\bibfnamefont {A.}~\bibnamefont {Weller}}, \bibinfo {author} {\bibfnamefont {S.}~\bibnamefont {Giovanazzi}},\ and\ \bibinfo {author} {\bibfnamefont {M.~K.}\ \bibnamefont {Oberthaler}},\ }\bibfield  {title} {\bibinfo {title} {{\textit{Squeezing and Entanglement in a Bose--Einstein Condensate}}},\ }\href@noop {} {\bibfield  {journal} {\bibinfo  {journal} {Nature}\ }\textbf {\bibinfo {volume} {455}},\ \bibinfo {pages} {1216} (\bibinfo {year} {2008})}\BibitemShut {NoStop}%
\bibitem [{\citenamefont {Islam}\ \emph {et~al.}(2015)\citenamefont {Islam}, \citenamefont {Ma}, \citenamefont {Preiss}, \citenamefont {Eric~Tai}, \citenamefont {Lukin}, \citenamefont {Rispoli},\ and\ \citenamefont {Greiner}}]{islam2015measuring}%
  \BibitemOpen
  \bibfield  {author} {\bibinfo {author} {\bibfnamefont {R.}~\bibnamefont {Islam}}, \bibinfo {author} {\bibfnamefont {R.}~\bibnamefont {Ma}}, \bibinfo {author} {\bibfnamefont {P.~M.}\ \bibnamefont {Preiss}}, \bibinfo {author} {\bibfnamefont {M.}~\bibnamefont {Eric~Tai}}, \bibinfo {author} {\bibfnamefont {A.}~\bibnamefont {Lukin}}, \bibinfo {author} {\bibfnamefont {M.}~\bibnamefont {Rispoli}},\ and\ \bibinfo {author} {\bibfnamefont {M.}~\bibnamefont {Greiner}},\ }\bibfield  {title} {\bibinfo {title} {{\textit{Measuring Entanglement Entropy in a Quantum Many-Body System}}},\ }\href@noop {} {\bibfield  {journal} {\bibinfo  {journal} {Nature}\ }\textbf {\bibinfo {volume} {528}},\ \bibinfo {pages} {77} (\bibinfo {year} {2015})}\BibitemShut {NoStop}%
\bibitem [{\citenamefont {Pezz{\'e}}\ and\ \citenamefont {Smerzi}(2009)}]{pezze2009entanglement}%
  \BibitemOpen
  \bibfield  {author} {\bibinfo {author} {\bibfnamefont {L.}~\bibnamefont {Pezz{\'e}}}\ and\ \bibinfo {author} {\bibfnamefont {A.}~\bibnamefont {Smerzi}},\ }\bibfield  {title} {\bibinfo {title} {{\textit{Entanglement, Nonlinear Dynamics, and the Heisenberg Limit}}},\ }\href@noop {} {\bibfield  {journal} {\bibinfo  {journal} {Phys. Rev. Lett.}\ }\textbf {\bibinfo {volume} {102}},\ \bibinfo {pages} {100401} (\bibinfo {year} {2009})}\BibitemShut {NoStop}%
\bibitem [{\citenamefont {Hyllus}\ \emph {et~al.}(2012)\citenamefont {Hyllus}, \citenamefont {Laskowski}, \citenamefont {Krischek}, \citenamefont {Schwemmer}, \citenamefont {Wieczorek}, \citenamefont {Weinfurter}, \citenamefont {Pezz{\'e}},\ and\ \citenamefont {Smerzi}}]{hyllus2012fisher}%
  \BibitemOpen
  \bibfield  {author} {\bibinfo {author} {\bibfnamefont {P.}~\bibnamefont {Hyllus}}, \bibinfo {author} {\bibfnamefont {W.}~\bibnamefont {Laskowski}}, \bibinfo {author} {\bibfnamefont {R.}~\bibnamefont {Krischek}}, \bibinfo {author} {\bibfnamefont {C.}~\bibnamefont {Schwemmer}}, \bibinfo {author} {\bibfnamefont {W.}~\bibnamefont {Wieczorek}}, \bibinfo {author} {\bibfnamefont {H.}~\bibnamefont {Weinfurter}}, \bibinfo {author} {\bibfnamefont {L.}~\bibnamefont {Pezz{\'e}}},\ and\ \bibinfo {author} {\bibfnamefont {A.}~\bibnamefont {Smerzi}},\ }\bibfield  {title} {\bibinfo {title} {{\textit{Fisher Information and Multiparticle Entanglement}}},\ }\href@noop {} {\bibfield  {journal} {\bibinfo  {journal} {Phys. Rev. A}\ }\textbf {\bibinfo {volume} {85}},\ \bibinfo {pages} {022321} (\bibinfo {year} {2012})}\BibitemShut {NoStop}%
\bibitem [{\citenamefont {T{\'o}th}(2012)}]{toth2012multipartite}%
  \BibitemOpen
  \bibfield  {author} {\bibinfo {author} {\bibfnamefont {G.}~\bibnamefont {T{\'o}th}},\ }\bibfield  {title} {\bibinfo {title} {{\textit{Multipartite Entanglement and High-Precision Metrology}}},\ }\href@noop {} {\bibfield  {journal} {\bibinfo  {journal} {Phys. Rev. A}\ }\textbf {\bibinfo {volume} {85}},\ \bibinfo {pages} {022322} (\bibinfo {year} {2012})}\BibitemShut {NoStop}%
\bibitem [{\citenamefont {Hauke}\ \emph {et~al.}(2016)\citenamefont {Hauke}, \citenamefont {Heyl}, \citenamefont {Tagliacozzo},\ and\ \citenamefont {Zoller}}]{hauke2016measuring}%
  \BibitemOpen
  \bibfield  {author} {\bibinfo {author} {\bibfnamefont {P.}~\bibnamefont {Hauke}}, \bibinfo {author} {\bibfnamefont {M.}~\bibnamefont {Heyl}}, \bibinfo {author} {\bibfnamefont {L.}~\bibnamefont {Tagliacozzo}},\ and\ \bibinfo {author} {\bibfnamefont {P.}~\bibnamefont {Zoller}},\ }\bibfield  {title} {\bibinfo {title} {{\textit{Measuring Multipartite Entanglement Through Dynamic Susceptibilities}}},\ }\href@noop {} {\bibfield  {journal} {\bibinfo  {journal} {Nat. Phys.}\ }\textbf {\bibinfo {volume} {12}},\ \bibinfo {pages} {778} (\bibinfo {year} {2016})}\BibitemShut {NoStop}%
\bibitem [{\citenamefont {Menon}\ \emph {et~al.}(2023)\citenamefont {Menon}, \citenamefont {Sherman}, \citenamefont {Dupont}, \citenamefont {Scheie}, \citenamefont {Tennant},\ and\ \citenamefont {Moore}}]{menon2023multipartite}%
  \BibitemOpen
  \bibfield  {author} {\bibinfo {author} {\bibfnamefont {V.}~\bibnamefont {Menon}}, \bibinfo {author} {\bibfnamefont {N.~E.}\ \bibnamefont {Sherman}}, \bibinfo {author} {\bibfnamefont {M.}~\bibnamefont {Dupont}}, \bibinfo {author} {\bibfnamefont {A.~O.}\ \bibnamefont {Scheie}}, \bibinfo {author} {\bibfnamefont {D.~A.}\ \bibnamefont {Tennant}},\ and\ \bibinfo {author} {\bibfnamefont {J.~E.}\ \bibnamefont {Moore}},\ }\bibfield  {title} {\bibinfo {title} {{\textit{Multipartite Entanglement in the One-Dimensional Spin$-1/2$ Heisenberg Antiferromagnet}}},\ }\href@noop {} {\bibfield  {journal} {\bibinfo  {journal} {Phys. Rev. B}\ }\textbf {\bibinfo {volume} {107}},\ \bibinfo {pages} {054422} (\bibinfo {year} {2023})}\BibitemShut {NoStop}%
\bibitem [{\citenamefont {Fang}\ \emph {et~al.}(2025)\citenamefont {Fang}, \citenamefont {Mahankali}, \citenamefont {Wang}, \citenamefont {Chen}, \citenamefont {Hu}, \citenamefont {Paschen},\ and\ \citenamefont {Si}}]{fang2025amplified}%
  \BibitemOpen
  \bibfield  {author} {\bibinfo {author} {\bibfnamefont {Y.}~\bibnamefont {Fang}}, \bibinfo {author} {\bibfnamefont {M.}~\bibnamefont {Mahankali}}, \bibinfo {author} {\bibfnamefont {Y.}~\bibnamefont {Wang}}, \bibinfo {author} {\bibfnamefont {L.}~\bibnamefont {Chen}}, \bibinfo {author} {\bibfnamefont {H.}~\bibnamefont {Hu}}, \bibinfo {author} {\bibfnamefont {S.}~\bibnamefont {Paschen}},\ and\ \bibinfo {author} {\bibfnamefont {Q.}~\bibnamefont {Si}},\ }\bibfield  {title} {\bibinfo {title} {\textit{Amplified multipartite entanglement witnessed in a quantum critical metal}},\ }\href@noop {} {\bibfield  {journal} {\bibinfo  {journal} {Nat. Commun.}\ }\textbf {\bibinfo {volume} {16}},\ \bibinfo {pages} {2498} (\bibinfo {year} {2025})}\BibitemShut {NoStop}%
\bibitem [{\citenamefont {Shen}\ \emph {et~al.}(2026)\citenamefont {Shen}, \citenamefont {Ding}, \citenamefont {Zhao}, \citenamefont {Evangelista},\ and\ \citenamefont {Wang}}]{shen2026witnessing}%
  \BibitemOpen
  \bibfield  {author} {\bibinfo {author} {\bibfnamefont {Z.}~\bibnamefont {Shen}}, \bibinfo {author} {\bibfnamefont {S.}~\bibnamefont {Ding}}, \bibinfo {author} {\bibfnamefont {Z.}~\bibnamefont {Zhao}}, \bibinfo {author} {\bibfnamefont {F.~A.}\ \bibnamefont {Evangelista}},\ and\ \bibinfo {author} {\bibfnamefont {Y.}~\bibnamefont {Wang}},\ }\bibfield  {title} {\bibinfo {title} {{\textit{Witnessing Spin-Orbital Entanglement Using Resonant Inelastic X-Ray Scattering}}},\ }\href@noop {} {\bibfield  {journal} {\bibinfo  {journal} {Phys. Rev. Lett.}\ }\textbf {\bibinfo {volume} {136}},\ \bibinfo {pages} {116504} (\bibinfo {year} {2026})}\BibitemShut {NoStop}%
\bibitem [{\citenamefont {Coffman}\ \emph {et~al.}(2000)\citenamefont {Coffman}, \citenamefont {Kundu},\ and\ \citenamefont {Wootters}}]{coffman2000distributed}%
  \BibitemOpen
  \bibfield  {author} {\bibinfo {author} {\bibfnamefont {V.}~\bibnamefont {Coffman}}, \bibinfo {author} {\bibfnamefont {J.}~\bibnamefont {Kundu}},\ and\ \bibinfo {author} {\bibfnamefont {W.~K.}\ \bibnamefont {Wootters}},\ }\bibfield  {title} {\bibinfo {title} {{\textit{Distributed Entanglement}}},\ }\href@noop {} {\bibfield  {journal} {\bibinfo  {journal} {Phys. Rev. A}\ }\textbf {\bibinfo {volume} {61}},\ \bibinfo {pages} {052306} (\bibinfo {year} {2000})}\BibitemShut {NoStop}%
\bibitem [{\citenamefont {Liu}\ \emph {et~al.}(2025)\citenamefont {Liu}, \citenamefont {Xu}, \citenamefont {Liu},\ and\ \citenamefont {Wang}}]{liu2025entanglement}%
  \BibitemOpen
  \bibfield  {author} {\bibinfo {author} {\bibfnamefont {T.}~\bibnamefont {Liu}}, \bibinfo {author} {\bibfnamefont {L.}~\bibnamefont {Xu}}, \bibinfo {author} {\bibfnamefont {J.}~\bibnamefont {Liu}},\ and\ \bibinfo {author} {\bibfnamefont {Y.}~\bibnamefont {Wang}},\ }\bibfield  {title} {\bibinfo {title} {{\textit{Entanglement Witness for Indistinguishable Electrons Using Solid-State Spectroscopy}}},\ }\href@noop {} {\bibfield  {journal} {\bibinfo  {journal} {Phys. Rev. X}\ }\textbf {\bibinfo {volume} {15}},\ \bibinfo {pages} {011056} (\bibinfo {year} {2025})}\BibitemShut {NoStop}%
\bibitem [{\citenamefont {Mathew}\ \emph {et~al.}(2020)\citenamefont {Mathew}, \citenamefont {Silva}, \citenamefont {Jain}, \citenamefont {Mohan}, \citenamefont {Adroja}, \citenamefont {Sakai}, \citenamefont {Tomy}, \citenamefont {Banerjee}, \citenamefont {Goreti}, \citenamefont {Aswathi}, \citenamefont {Singh},\ and\ \citenamefont {Jaiswal-Nagar}}]{mathew2020experimental}%
  \BibitemOpen
  \bibfield  {author} {\bibinfo {author} {\bibfnamefont {G.}~\bibnamefont {Mathew}}, \bibinfo {author} {\bibfnamefont {S.~L.~L.}\ \bibnamefont {Silva}}, \bibinfo {author} {\bibfnamefont {A.}~\bibnamefont {Jain}}, \bibinfo {author} {\bibfnamefont {A.}~\bibnamefont {Mohan}}, \bibinfo {author} {\bibfnamefont {D.~T.}\ \bibnamefont {Adroja}}, \bibinfo {author} {\bibfnamefont {V.~G.}\ \bibnamefont {Sakai}}, \bibinfo {author} {\bibfnamefont {C.~V.}\ \bibnamefont {Tomy}}, \bibinfo {author} {\bibfnamefont {A.}~\bibnamefont {Banerjee}}, \bibinfo {author} {\bibfnamefont {R.}~\bibnamefont {Goreti}}, \bibinfo {author} {\bibfnamefont {V.~N.}\ \bibnamefont {Aswathi}}, \bibinfo {author} {\bibfnamefont {R.}~\bibnamefont {Singh}},\ and\ \bibinfo {author} {\bibfnamefont {D.}~\bibnamefont {Jaiswal-Nagar}},\ }\bibfield  {title} {\bibinfo {title} {{\textit{Experimental Realisation of Multipartite Entanglement Via Quantum Fisher Information in a Uniform Antiferromagnetic Quantum Spin Chain}}},\ }\href@noop {} {\bibfield  {journal}
  {\bibinfo  {journal} {Phys. Rev. Research}\ }\textbf {\bibinfo {volume} {2}},\ \bibinfo {pages} {043329} (\bibinfo {year} {2020})}\BibitemShut {NoStop}%
\bibitem [{\citenamefont {Scheie}\ \emph {et~al.}(2021)\citenamefont {Scheie}, \citenamefont {Laurell}, \citenamefont {Samarakoon}, \citenamefont {Lake}, \citenamefont {Nagler}, \citenamefont {Granroth}, \citenamefont {Okamoto}, \citenamefont {Alvarez},\ and\ \citenamefont {Tennant}}]{scheie2021witnessing}%
  \BibitemOpen
  \bibfield  {author} {\bibinfo {author} {\bibfnamefont {A.}~\bibnamefont {Scheie}}, \bibinfo {author} {\bibfnamefont {P.}~\bibnamefont {Laurell}}, \bibinfo {author} {\bibfnamefont {A.~M.}\ \bibnamefont {Samarakoon}}, \bibinfo {author} {\bibfnamefont {B.}~\bibnamefont {Lake}}, \bibinfo {author} {\bibfnamefont {S.~E.}\ \bibnamefont {Nagler}}, \bibinfo {author} {\bibfnamefont {G.~E.}\ \bibnamefont {Granroth}}, \bibinfo {author} {\bibfnamefont {S.}~\bibnamefont {Okamoto}}, \bibinfo {author} {\bibfnamefont {G.}~\bibnamefont {Alvarez}},\ and\ \bibinfo {author} {\bibfnamefont {D.~A.}\ \bibnamefont {Tennant}},\ }\bibfield  {title} {\bibinfo {title} {{\textit{Witnessing Entanglement in Quantum Magnets Using Neutron Scattering}}},\ }\href@noop {} {\bibfield  {journal} {\bibinfo  {journal} {Phys. Rev. B}\ }\textbf {\bibinfo {volume} {103}},\ \bibinfo {pages} {224434} (\bibinfo {year} {2021})}\BibitemShut {NoStop}%
\bibitem [{\citenamefont {Laurell}\ \emph {et~al.}(2021)\citenamefont {Laurell}, \citenamefont {Scheie}, \citenamefont {Mukherjee}, \citenamefont {Koza}, \citenamefont {Enderle}, \citenamefont {Tylczynski}, \citenamefont {Okamoto}, \citenamefont {Coldea}, \citenamefont {Tennant},\ and\ \citenamefont {Alvarez}}]{laurell2021quantifying}%
  \BibitemOpen
  \bibfield  {author} {\bibinfo {author} {\bibfnamefont {P.}~\bibnamefont {Laurell}}, \bibinfo {author} {\bibfnamefont {A.}~\bibnamefont {Scheie}}, \bibinfo {author} {\bibfnamefont {C.~J.}\ \bibnamefont {Mukherjee}}, \bibinfo {author} {\bibfnamefont {M.~M.}\ \bibnamefont {Koza}}, \bibinfo {author} {\bibfnamefont {M.}~\bibnamefont {Enderle}}, \bibinfo {author} {\bibfnamefont {Z.}~\bibnamefont {Tylczynski}}, \bibinfo {author} {\bibfnamefont {S.}~\bibnamefont {Okamoto}}, \bibinfo {author} {\bibfnamefont {R.}~\bibnamefont {Coldea}}, \bibinfo {author} {\bibfnamefont {D.~A.}\ \bibnamefont {Tennant}},\ and\ \bibinfo {author} {\bibfnamefont {G.}~\bibnamefont {Alvarez}},\ }\bibfield  {title} {\bibinfo {title} {{\textit{Quantifying and Controlling Entanglement in the Quantum Magnet Cs$_2$CoCl$_4$}}},\ }\href@noop {} {\bibfield  {journal} {\bibinfo  {journal} {Phys. Rev. Lett.}\ }\textbf {\bibinfo {volume} {127}},\ \bibinfo {pages} {037201} (\bibinfo {year} {2021})}\BibitemShut {NoStop}%
\bibitem [{\citenamefont {Scheie}\ \emph {et~al.}(2024)\citenamefont {Scheie}, \citenamefont {Ghioldi}, \citenamefont {Xing}, \citenamefont {Paddison}, \citenamefont {Sherman}, \citenamefont {Dupont}, \citenamefont {Sanjeewa}, \citenamefont {Lee}, \citenamefont {Woods}, \citenamefont {Abernathy} \emph {et~al.}}]{scheie2024proximate}%
  \BibitemOpen
  \bibfield  {author} {\bibinfo {author} {\bibfnamefont {A.}~\bibnamefont {Scheie}}, \bibinfo {author} {\bibfnamefont {E.}~\bibnamefont {Ghioldi}}, \bibinfo {author} {\bibfnamefont {J.}~\bibnamefont {Xing}}, \bibinfo {author} {\bibfnamefont {J.}~\bibnamefont {Paddison}}, \bibinfo {author} {\bibfnamefont {N.}~\bibnamefont {Sherman}}, \bibinfo {author} {\bibfnamefont {M.}~\bibnamefont {Dupont}}, \bibinfo {author} {\bibfnamefont {L.}~\bibnamefont {Sanjeewa}}, \bibinfo {author} {\bibfnamefont {S.}~\bibnamefont {Lee}}, \bibinfo {author} {\bibfnamefont {A.}~\bibnamefont {Woods}}, \bibinfo {author} {\bibfnamefont {D.}~\bibnamefont {Abernathy}}, \emph {et~al.},\ }\bibfield  {title} {\bibinfo {title} {{\textit{Proximate Spin Liquid and Fractionalization in the Triangular Antiferromagnet KYbSe$_2$}}},\ }\href@noop {} {\bibfield  {journal} {\bibinfo  {journal} {Nat. Phys.}\ }\textbf {\bibinfo {volume} {20}},\ \bibinfo {pages} {74} (\bibinfo {year} {2024})}\BibitemShut {NoStop}%
\bibitem [{\citenamefont {Mazza}\ \emph {et~al.}(2026)\citenamefont {Mazza}, \citenamefont {Biswas}, \citenamefont {Yan}, \citenamefont {Prokofiev}, \citenamefont {Steffens}, \citenamefont {Si}, \citenamefont {Assaad},\ and\ \citenamefont {Paschen}}]{mazza2026quantum}%
  \BibitemOpen
  \bibfield  {author} {\bibinfo {author} {\bibfnamefont {F.}~\bibnamefont {Mazza}}, \bibinfo {author} {\bibfnamefont {S.}~\bibnamefont {Biswas}}, \bibinfo {author} {\bibfnamefont {X.}~\bibnamefont {Yan}}, \bibinfo {author} {\bibfnamefont {A.}~\bibnamefont {Prokofiev}}, \bibinfo {author} {\bibfnamefont {P.}~\bibnamefont {Steffens}}, \bibinfo {author} {\bibfnamefont {Q.}~\bibnamefont {Si}}, \bibinfo {author} {\bibfnamefont {F.~F.}\ \bibnamefont {Assaad}},\ and\ \bibinfo {author} {\bibfnamefont {S.}~\bibnamefont {Paschen}},\ }\bibfield  {title} {\bibinfo {title} {{\textit{Quantum Fisher Information in a Strange Metal}}},\ }\href@noop {} {\bibfield  {journal} {\bibinfo  {journal} {Nat. Phys.}\ }\textbf {\bibinfo {volume} {22}},\ \bibinfo {pages} {1064} (\bibinfo {year} {2026})}\BibitemShut {NoStop}%
\bibitem [{\citenamefont {de~Picciotto}\ \emph {et~al.}(1997)\citenamefont {de~Picciotto}, \citenamefont {Reznikov}, \citenamefont {Heiblum}, \citenamefont {Umansky}, \citenamefont {Bunin},\ and\ \citenamefont {Mahalu}}]{depicciotto1997direct}%
  \BibitemOpen
  \bibfield  {author} {\bibinfo {author} {\bibfnamefont {R.}~\bibnamefont {de~Picciotto}}, \bibinfo {author} {\bibfnamefont {M.}~\bibnamefont {Reznikov}}, \bibinfo {author} {\bibfnamefont {M.}~\bibnamefont {Heiblum}}, \bibinfo {author} {\bibfnamefont {V.}~\bibnamefont {Umansky}}, \bibinfo {author} {\bibfnamefont {G.}~\bibnamefont {Bunin}},\ and\ \bibinfo {author} {\bibfnamefont {D.}~\bibnamefont {Mahalu}},\ }\bibfield  {title} {\bibinfo {title} {{\textit{Direct Observation of a Fractional Charge}}},\ }\href@noop {} {\bibfield  {journal} {\bibinfo  {journal} {Nature}\ }\textbf {\bibinfo {volume} {389}},\ \bibinfo {pages} {162} (\bibinfo {year} {1997})}\BibitemShut {NoStop}%
\bibitem [{\citenamefont {Dolev}\ \emph {et~al.}(2008)\citenamefont {Dolev}, \citenamefont {Heiblum}, \citenamefont {Umansky}, \citenamefont {Stern},\ and\ \citenamefont {Mahalu}}]{dolev2008observation}%
  \BibitemOpen
  \bibfield  {author} {\bibinfo {author} {\bibfnamefont {M.}~\bibnamefont {Dolev}}, \bibinfo {author} {\bibfnamefont {M.}~\bibnamefont {Heiblum}}, \bibinfo {author} {\bibfnamefont {V.}~\bibnamefont {Umansky}}, \bibinfo {author} {\bibfnamefont {A.}~\bibnamefont {Stern}},\ and\ \bibinfo {author} {\bibfnamefont {D.}~\bibnamefont {Mahalu}},\ }\bibfield  {title} {\bibinfo {title} {{\textit{Observation of a Quarter of an Electron Charge at the $\nu = $5/2 Quantum Hall State}}},\ }\href@noop {} {\bibfield  {journal} {\bibinfo  {journal} {Nature}\ }\textbf {\bibinfo {volume} {452}},\ \bibinfo {pages} {829} (\bibinfo {year} {2008})}\BibitemShut {NoStop}%
\bibitem [{\citenamefont {Ghirardi}\ \emph {et~al.}(2002)\citenamefont {Ghirardi}, \citenamefont {Marinatto},\ and\ \citenamefont {Weber}}]{ghirardi2002entanglement}%
  \BibitemOpen
  \bibfield  {author} {\bibinfo {author} {\bibfnamefont {G.}~\bibnamefont {Ghirardi}}, \bibinfo {author} {\bibfnamefont {L.}~\bibnamefont {Marinatto}},\ and\ \bibinfo {author} {\bibfnamefont {T.}~\bibnamefont {Weber}},\ }\bibfield  {title} {\bibinfo {title} {{\textit{Entanglement and Properties of Composite Quantum Systems: A Conceptual and Mathematical Analysis}}},\ }\href@noop {} {\bibfield  {journal} {\bibinfo  {journal} {J. Stat. Phys.}\ }\textbf {\bibinfo {volume} {108}},\ \bibinfo {pages} {49} (\bibinfo {year} {2002})}\BibitemShut {NoStop}%
\bibitem [{\citenamefont {Ghirardi}\ and\ \citenamefont {Marinatto}(2004)}]{ghirardi2004general}%
  \BibitemOpen
  \bibfield  {author} {\bibinfo {author} {\bibfnamefont {G.}~\bibnamefont {Ghirardi}}\ and\ \bibinfo {author} {\bibfnamefont {L.}~\bibnamefont {Marinatto}},\ }\bibfield  {title} {\bibinfo {title} {{\textit{General Criterion for the Entanglement of Two Indistinguishable Particles}}},\ }\href@noop {} {\bibfield  {journal} {\bibinfo  {journal} {Phys. Rev. A}\ }\textbf {\bibinfo {volume} {70}},\ \bibinfo {pages} {012109} (\bibinfo {year} {2004})}\BibitemShut {NoStop}%
\bibitem [{\citenamefont {Amico}\ \emph {et~al.}(2008)\citenamefont {Amico}, \citenamefont {Fazio}, \citenamefont {Osterloh},\ and\ \citenamefont {Vedral}}]{amico2008entanglement}%
  \BibitemOpen
  \bibfield  {author} {\bibinfo {author} {\bibfnamefont {L.}~\bibnamefont {Amico}}, \bibinfo {author} {\bibfnamefont {R.}~\bibnamefont {Fazio}}, \bibinfo {author} {\bibfnamefont {A.}~\bibnamefont {Osterloh}},\ and\ \bibinfo {author} {\bibfnamefont {V.}~\bibnamefont {Vedral}},\ }\bibfield  {title} {\bibinfo {title} {{\textit{Entanglement in Many-Body Systems}}},\ }\href@noop {} {\bibfield  {journal} {\bibinfo  {journal} {Rev. Mod. Phys.}\ }\textbf {\bibinfo {volume} {80}},\ \bibinfo {pages} {517} (\bibinfo {year} {2008})}\BibitemShut {NoStop}%
\bibitem [{\citenamefont {Costa~de Almeida}\ and\ \citenamefont {Hauke}(2021)}]{almeida2021from}%
  \BibitemOpen
  \bibfield  {author} {\bibinfo {author} {\bibfnamefont {R.}~\bibnamefont {Costa~de Almeida}}\ and\ \bibinfo {author} {\bibfnamefont {P.}~\bibnamefont {Hauke}},\ }\bibfield  {title} {\bibinfo {title} {{\textit{From Entanglement Certification with Quench Dynamics to Multipartite Entanglement of Interacting Fermions}}},\ }\href@noop {} {\bibfield  {journal} {\bibinfo  {journal} {Phys. Rev. Research}\ }\textbf {\bibinfo {volume} {3}},\ \bibinfo {pages} {L032051} (\bibinfo {year} {2021})}\BibitemShut {NoStop}%
\bibitem [{\citenamefont {Blanter}\ and\ \citenamefont {B{\"u}ttiker}(2000)}]{blanter2000shot}%
  \BibitemOpen
  \bibfield  {author} {\bibinfo {author} {\bibfnamefont {Y.~M.}\ \bibnamefont {Blanter}}\ and\ \bibinfo {author} {\bibfnamefont {M.}~\bibnamefont {B{\"u}ttiker}},\ }\bibfield  {title} {\bibinfo {title} {{\textit{Shot Noise in Mesoscopic Conductors}}},\ }\href@noop {} {\bibfield  {journal} {\bibinfo  {journal} {Phys. Rep.}\ }\textbf {\bibinfo {volume} {336}},\ \bibinfo {pages} {1} (\bibinfo {year} {2000})}\BibitemShut {NoStop}%
\bibitem [{\citenamefont {Clerk}\ \emph {et~al.}(2010)\citenamefont {Clerk}, \citenamefont {Devoret}, \citenamefont {Girvin}, \citenamefont {Marquardt},\ and\ \citenamefont {Schoelkopf}}]{clerk2010introduction}%
  \BibitemOpen
  \bibfield  {author} {\bibinfo {author} {\bibfnamefont {A.~A.}\ \bibnamefont {Clerk}}, \bibinfo {author} {\bibfnamefont {M.~H.}\ \bibnamefont {Devoret}}, \bibinfo {author} {\bibfnamefont {S.~M.}\ \bibnamefont {Girvin}}, \bibinfo {author} {\bibfnamefont {F.}~\bibnamefont {Marquardt}},\ and\ \bibinfo {author} {\bibfnamefont {R.~J.}\ \bibnamefont {Schoelkopf}},\ }\bibfield  {title} {\bibinfo {title} {{\textit{Introduction to Quantum Noise, Measurement, and Amplification}}},\ }\href@noop {} {\bibfield  {journal} {\bibinfo  {journal} {Rev. Mod. Phys.}\ }\textbf {\bibinfo {volume} {82}},\ \bibinfo {pages} {1155} (\bibinfo {year} {2010})}\BibitemShut {NoStop}%
\bibitem [{\citenamefont {Schoelkopf}\ \emph {et~al.}(1997)\citenamefont {Schoelkopf}, \citenamefont {Burke}, \citenamefont {Kozhevnikov}, \citenamefont {Prober},\ and\ \citenamefont {Rooks}}]{schoelkopf1997frequency}%
  \BibitemOpen
  \bibfield  {author} {\bibinfo {author} {\bibfnamefont {R.}~\bibnamefont {Schoelkopf}}, \bibinfo {author} {\bibfnamefont {P.}~\bibnamefont {Burke}}, \bibinfo {author} {\bibfnamefont {A.}~\bibnamefont {Kozhevnikov}}, \bibinfo {author} {\bibfnamefont {D.}~\bibnamefont {Prober}},\ and\ \bibinfo {author} {\bibfnamefont {M.}~\bibnamefont {Rooks}},\ }\bibfield  {title} {\bibinfo {title} {{\textit{Frequency Dependence of Shot Noise in a Diffusive Mesoscopic Conductor}}},\ }\href@noop {} {\bibfield  {journal} {\bibinfo  {journal} {Phys. Rev. Lett.}\ }\textbf {\bibinfo {volume} {78}},\ \bibinfo {pages} {3370} (\bibinfo {year} {1997})}\BibitemShut {NoStop}%
\bibitem [{\citenamefont {Zakka-Bajjani}\ \emph {et~al.}(2007)\citenamefont {Zakka-Bajjani}, \citenamefont {Segala}, \citenamefont {Portier}, \citenamefont {Roche}, \citenamefont {Glattli}, \citenamefont {Cavanna},\ and\ \citenamefont {Jin}}]{zakka2007experimental}%
  \BibitemOpen
  \bibfield  {author} {\bibinfo {author} {\bibfnamefont {E.}~\bibnamefont {Zakka-Bajjani}}, \bibinfo {author} {\bibfnamefont {J.}~\bibnamefont {Segala}}, \bibinfo {author} {\bibfnamefont {F.}~\bibnamefont {Portier}}, \bibinfo {author} {\bibfnamefont {P.}~\bibnamefont {Roche}}, \bibinfo {author} {\bibfnamefont {D.}~\bibnamefont {Glattli}}, \bibinfo {author} {\bibfnamefont {A.}~\bibnamefont {Cavanna}},\ and\ \bibinfo {author} {\bibfnamefont {Y.}~\bibnamefont {Jin}},\ }\bibfield  {title} {\bibinfo {title} {{\textit{Experimental Test of the High-Frequency Quantum Shot Noise Theory in a Quantum Point Contact}}},\ }\href@noop {} {\bibfield  {journal} {\bibinfo  {journal} {Phys. Rev. Lett.}\ }\textbf {\bibinfo {volume} {99}},\ \bibinfo {pages} {236803} (\bibinfo {year} {2007})}\BibitemShut {NoStop}%
\bibitem [{\citenamefont {Kubo}(1957)}]{kubo1957statistical}%
  \BibitemOpen
  \bibfield  {author} {\bibinfo {author} {\bibfnamefont {R.}~\bibnamefont {Kubo}},\ }\bibfield  {title} {\bibinfo {title} {{\textit{Statistical-Mechanical Theory of Irreversible Processes. {I}. General Theory and Simple Applications to Magnetic and Conduction Problems}}},\ }\href@noop {} {\bibfield  {journal} {\bibinfo  {journal} {J. Phys. Soc. Jpn.}\ }\textbf {\bibinfo {volume} {12}},\ \bibinfo {pages} {570} (\bibinfo {year} {1957})}\BibitemShut {NoStop}%
\bibitem [{\citenamefont {Klich}\ and\ \citenamefont {Levitov}(2009)}]{klich2009quantum}%
  \BibitemOpen
  \bibfield  {author} {\bibinfo {author} {\bibfnamefont {I.}~\bibnamefont {Klich}}\ and\ \bibinfo {author} {\bibfnamefont {L.}~\bibnamefont {Levitov}},\ }\bibfield  {title} {\bibinfo {title} {{\textit{Quantum Noise as an Entanglement Meter}}},\ }\href@noop {} {\bibfield  {journal} {\bibinfo  {journal} {Phys. Rev. Lett.}\ }\textbf {\bibinfo {volume} {102}},\ \bibinfo {pages} {100502} (\bibinfo {year} {2009})}\BibitemShut {NoStop}%
\bibitem [{\citenamefont {Johnson}(1928)}]{johnson1928thermal}%
  \BibitemOpen
  \bibfield  {author} {\bibinfo {author} {\bibfnamefont {J.~B.}\ \bibnamefont {Johnson}},\ }\bibfield  {title} {\bibinfo {title} {{\textit{Thermal Agitation of Electricity in Conductors}}},\ }\href@noop {} {\bibfield  {journal} {\bibinfo  {journal} {Phys. Rev.}\ }\textbf {\bibinfo {volume} {32}},\ \bibinfo {pages} {97} (\bibinfo {year} {1928})}\BibitemShut {NoStop}%
\bibitem [{\citenamefont {Nyquist}(1928)}]{nyquist1928thermal}%
  \BibitemOpen
  \bibfield  {author} {\bibinfo {author} {\bibfnamefont {H.}~\bibnamefont {Nyquist}},\ }\bibfield  {title} {\bibinfo {title} {{\textit{Thermal Agitation of Electric Charge in Conductors}}},\ }\href@noop {} {\bibfield  {journal} {\bibinfo  {journal} {Phys. Rev.}\ }\textbf {\bibinfo {volume} {32}},\ \bibinfo {pages} {110} (\bibinfo {year} {1928})}\BibitemShut {NoStop}%
\bibitem [{\citenamefont {Jarrell}\ and\ \citenamefont {Gubernatis}(1996)}]{JarrellGubernatis1996}%
  \BibitemOpen
  \bibfield  {author} {\bibinfo {author} {\bibfnamefont {M.}~\bibnamefont {Jarrell}}\ and\ \bibinfo {author} {\bibfnamefont {J.~E.}\ \bibnamefont {Gubernatis}},\ }\bibfield  {title} {\bibinfo {title} {{\textit{Bayesian Inference and the Analytic Continuation of Imaginary-Time Quantum Monte Carlo Data}}},\ }\href@noop {} {\bibfield  {journal} {\bibinfo  {journal} {Plasma Phys. Rep.}\ }\textbf {\bibinfo {volume} {269}},\ \bibinfo {pages} {133} (\bibinfo {year} {1996})}\BibitemShut {NoStop}%
\bibitem [{\citenamefont {Gunnarsson}\ \emph {et~al.}(2010)\citenamefont {Gunnarsson}, \citenamefont {Haverkort},\ and\ \citenamefont {Sangiovanni}}]{GunnarssonEtAl2010}%
  \BibitemOpen
  \bibfield  {author} {\bibinfo {author} {\bibfnamefont {O.}~\bibnamefont {Gunnarsson}}, \bibinfo {author} {\bibfnamefont {M.~W.}\ \bibnamefont {Haverkort}},\ and\ \bibinfo {author} {\bibfnamefont {G.}~\bibnamefont {Sangiovanni}},\ }\bibfield  {title} {\bibinfo {title} {{\textit{Analytical Continuation of Imaginary Axis Data for Optical Spectroscopy}}},\ }\href@noop {} {\bibfield  {journal} {\bibinfo  {journal} {Phys. Rev. B}\ }\textbf {\bibinfo {volume} {82}},\ \bibinfo {pages} {165125} (\bibinfo {year} {2010})}\BibitemShut {NoStop}%
\bibitem [{\citenamefont {Juh{\'a}sz}\ and\ \citenamefont {Mazziotti}(2006)}]{juhasz2006cumulant}%
  \BibitemOpen
  \bibfield  {author} {\bibinfo {author} {\bibfnamefont {T.}~\bibnamefont {Juh{\'a}sz}}\ and\ \bibinfo {author} {\bibfnamefont {D.~A.}\ \bibnamefont {Mazziotti}},\ }\bibfield  {title} {\bibinfo {title} {{\textit{The Cumulant Two-Particle Reduced Density Matrix as a Measure of Electron Correlation and Entanglement}}},\ }\href@noop {} {\bibfield  {journal} {\bibinfo  {journal} {The Journal of Chemical Physics}\ }\textbf {\bibinfo {volume} {125}} (\bibinfo {year} {2006})}\BibitemShut {NoStop}%
\bibitem [{\citenamefont {Schouten}\ \emph {et~al.}(2022)\citenamefont {Schouten}, \citenamefont {Sager-Smith},\ and\ \citenamefont {Mazziotti}}]{schouten2022large}%
  \BibitemOpen
  \bibfield  {author} {\bibinfo {author} {\bibfnamefont {A.~O.}\ \bibnamefont {Schouten}}, \bibinfo {author} {\bibfnamefont {L.~M.}\ \bibnamefont {Sager-Smith}},\ and\ \bibinfo {author} {\bibfnamefont {D.~A.}\ \bibnamefont {Mazziotti}},\ }\bibfield  {title} {\bibinfo {title} {{\textit{Large Cumulant Eigenvalue as a Signature of Exciton Condensation}}},\ }\href@noop {} {\bibfield  {journal} {\bibinfo  {journal} {Phys. Rev. B}\ }\textbf {\bibinfo {volume} {105}},\ \bibinfo {pages} {245151} (\bibinfo {year} {2022})}\BibitemShut {NoStop}%
\bibitem [{ent()}]{entanglementCondition}%
  \BibitemOpen
  \href@noop {} {}\bibinfo {note} {The threshold crossing is a sufficient test, not a necessary condition for entanglement. Exceeding $\mu(k)$ excludes $k$-producible states, whereas remaining below it does not establish separability or an upper bound on entanglement depth. Thus, the onset near $U=6t_h$ marks the sensitivity of this specific witness rather than the onset of entanglement in the Hubbard ground state.}\BibitemShut {Stop}%
\bibitem [{\citenamefont {Devakul}\ \emph {et~al.}(2021)\citenamefont {Devakul}, \citenamefont {Cr{\'e}pel}, \citenamefont {Zhang},\ and\ \citenamefont {Fu}}]{devakul2021magic}%
  \BibitemOpen
  \bibfield  {author} {\bibinfo {author} {\bibfnamefont {T.}~\bibnamefont {Devakul}}, \bibinfo {author} {\bibfnamefont {V.}~\bibnamefont {Cr{\'e}pel}}, \bibinfo {author} {\bibfnamefont {Y.}~\bibnamefont {Zhang}},\ and\ \bibinfo {author} {\bibfnamefont {L.}~\bibnamefont {Fu}},\ }\bibfield  {title} {\bibinfo {title} {{\textit{Magic in Twisted Transition Metal Dichalcogenide Bilayers}}},\ }\href@noop {} {\bibfield  {journal} {\bibinfo  {journal} {Nat. Commun.}\ }\textbf {\bibinfo {volume} {12}},\ \bibinfo {pages} {6730} (\bibinfo {year} {2021})}\BibitemShut {NoStop}%
\bibitem [{\citenamefont {Li}\ \emph {et~al.}(2021)\citenamefont {Li}, \citenamefont {Kumar}, \citenamefont {Sun},\ and\ \citenamefont {Lin}}]{li2021spontaneous}%
  \BibitemOpen
  \bibfield  {author} {\bibinfo {author} {\bibfnamefont {H.}~\bibnamefont {Li}}, \bibinfo {author} {\bibfnamefont {U.}~\bibnamefont {Kumar}}, \bibinfo {author} {\bibfnamefont {K.}~\bibnamefont {Sun}},\ and\ \bibinfo {author} {\bibfnamefont {S.-Z.}\ \bibnamefont {Lin}},\ }\bibfield  {title} {\bibinfo {title} {{\textit{Spontaneous Fractional Chern Insulators in Transition Metal Dichalcogenide Moir{\'e} Superlattices}}},\ }\href@noop {} {\bibfield  {journal} {\bibinfo  {journal} {Phys. Rev. Research}\ }\textbf {\bibinfo {volume} {3}},\ \bibinfo {pages} {L032070} (\bibinfo {year} {2021})}\BibitemShut {NoStop}%
\bibitem [{\citenamefont {Cai}\ \emph {et~al.}(2023)\citenamefont {Cai}, \citenamefont {Anderson}, \citenamefont {Wang}, \citenamefont {Zhang}, \citenamefont {Liu}, \citenamefont {Holtzmann}, \citenamefont {Zhang}, \citenamefont {Fan}, \citenamefont {Taniguchi}, \citenamefont {Watanabe} \emph {et~al.}}]{cai2023signatures}%
  \BibitemOpen
  \bibfield  {author} {\bibinfo {author} {\bibfnamefont {J.}~\bibnamefont {Cai}}, \bibinfo {author} {\bibfnamefont {E.}~\bibnamefont {Anderson}}, \bibinfo {author} {\bibfnamefont {C.}~\bibnamefont {Wang}}, \bibinfo {author} {\bibfnamefont {X.}~\bibnamefont {Zhang}}, \bibinfo {author} {\bibfnamefont {X.}~\bibnamefont {Liu}}, \bibinfo {author} {\bibfnamefont {W.}~\bibnamefont {Holtzmann}}, \bibinfo {author} {\bibfnamefont {Y.}~\bibnamefont {Zhang}}, \bibinfo {author} {\bibfnamefont {F.}~\bibnamefont {Fan}}, \bibinfo {author} {\bibfnamefont {T.}~\bibnamefont {Taniguchi}}, \bibinfo {author} {\bibfnamefont {K.}~\bibnamefont {Watanabe}}, \emph {et~al.},\ }\bibfield  {title} {\bibinfo {title} {{\textit{Signatures of Fractional Quantum Anomalous Hall States in Twisted MoTe$_2$}}},\ }\href@noop {} {\bibfield  {journal} {\bibinfo  {journal} {Nature}\ }\textbf {\bibinfo {volume} {622}},\ \bibinfo {pages} {63} (\bibinfo {year} {2023})}\BibitemShut {NoStop}%
\bibitem [{\citenamefont {Zeng}\ \emph {et~al.}(2023)\citenamefont {Zeng}, \citenamefont {Xia}, \citenamefont {Kang}, \citenamefont {Zhu}, \citenamefont {Kn{\"u}ppel}, \citenamefont {Vaswani}, \citenamefont {Watanabe}, \citenamefont {Taniguchi}, \citenamefont {Mak},\ and\ \citenamefont {Shan}}]{zeng2023thermodynamic}%
  \BibitemOpen
  \bibfield  {author} {\bibinfo {author} {\bibfnamefont {Y.}~\bibnamefont {Zeng}}, \bibinfo {author} {\bibfnamefont {Z.}~\bibnamefont {Xia}}, \bibinfo {author} {\bibfnamefont {K.}~\bibnamefont {Kang}}, \bibinfo {author} {\bibfnamefont {J.}~\bibnamefont {Zhu}}, \bibinfo {author} {\bibfnamefont {P.}~\bibnamefont {Kn{\"u}ppel}}, \bibinfo {author} {\bibfnamefont {C.}~\bibnamefont {Vaswani}}, \bibinfo {author} {\bibfnamefont {K.}~\bibnamefont {Watanabe}}, \bibinfo {author} {\bibfnamefont {T.}~\bibnamefont {Taniguchi}}, \bibinfo {author} {\bibfnamefont {K.~F.}\ \bibnamefont {Mak}},\ and\ \bibinfo {author} {\bibfnamefont {J.}~\bibnamefont {Shan}},\ }\bibfield  {title} {\bibinfo {title} {\textit{Thermodynamic evidence of fractional Chern insulator in moir{\'e} MoTe2}},\ }\href@noop {} {\bibfield  {journal} {\bibinfo  {journal} {Nature}\ }\textbf {\bibinfo {volume} {622}},\ \bibinfo {pages} {69} (\bibinfo {year} {2023})}\BibitemShut {NoStop}%
\bibitem [{\citenamefont {Park}\ \emph {et~al.}(2023)\citenamefont {Park}, \citenamefont {Cai}, \citenamefont {Anderson}, \citenamefont {Zhang}, \citenamefont {Zhu}, \citenamefont {Liu}, \citenamefont {Wang}, \citenamefont {Holtzmann}, \citenamefont {Hu}, \citenamefont {Liu} \emph {et~al.}}]{park2023observation}%
  \BibitemOpen
  \bibfield  {author} {\bibinfo {author} {\bibfnamefont {H.}~\bibnamefont {Park}}, \bibinfo {author} {\bibfnamefont {J.}~\bibnamefont {Cai}}, \bibinfo {author} {\bibfnamefont {E.}~\bibnamefont {Anderson}}, \bibinfo {author} {\bibfnamefont {Y.}~\bibnamefont {Zhang}}, \bibinfo {author} {\bibfnamefont {J.}~\bibnamefont {Zhu}}, \bibinfo {author} {\bibfnamefont {X.}~\bibnamefont {Liu}}, \bibinfo {author} {\bibfnamefont {C.}~\bibnamefont {Wang}}, \bibinfo {author} {\bibfnamefont {W.}~\bibnamefont {Holtzmann}}, \bibinfo {author} {\bibfnamefont {C.}~\bibnamefont {Hu}}, \bibinfo {author} {\bibfnamefont {Z.}~\bibnamefont {Liu}}, \emph {et~al.},\ }\bibfield  {title} {\bibinfo {title} {\textit{Observation of fractionally quantized anomalous Hall effect}},\ }\href@noop {} {\bibfield  {journal} {\bibinfo  {journal} {Nature}\ }\textbf {\bibinfo {volume} {622}},\ \bibinfo {pages} {74} (\bibinfo {year} {2023})}\BibitemShut {NoStop}%
\bibitem [{\citenamefont {Li}\ and\ \citenamefont {Haldane}(2008)}]{li2008entanglement}%
  \BibitemOpen
  \bibfield  {author} {\bibinfo {author} {\bibfnamefont {H.}~\bibnamefont {Li}}\ and\ \bibinfo {author} {\bibfnamefont {F.~D.~M.}\ \bibnamefont {Haldane}},\ }\bibfield  {title} {\bibinfo {title} {{\textit{Entanglement Spectrum as a Generalization of Entanglement Entropy: Identification of Topological Order in Non-Abelian Fractional Quantum Hall Effect States}}},\ }\href@noop {} {\bibfield  {journal} {\bibinfo  {journal} {Phys. Rev. Lett.}\ }\textbf {\bibinfo {volume} {101}},\ \bibinfo {pages} {010504} (\bibinfo {year} {2008})}\BibitemShut {NoStop}%
\bibitem [{\citenamefont {Reddy}\ \emph {et~al.}(2023)\citenamefont {Reddy}, \citenamefont {Alsallom}, \citenamefont {Zhang}, \citenamefont {Devakul},\ and\ \citenamefont {Fu}}]{reddy2023fractional}%
  \BibitemOpen
  \bibfield  {author} {\bibinfo {author} {\bibfnamefont {A.~P.}\ \bibnamefont {Reddy}}, \bibinfo {author} {\bibfnamefont {F.}~\bibnamefont {Alsallom}}, \bibinfo {author} {\bibfnamefont {Y.}~\bibnamefont {Zhang}}, \bibinfo {author} {\bibfnamefont {T.}~\bibnamefont {Devakul}},\ and\ \bibinfo {author} {\bibfnamefont {L.}~\bibnamefont {Fu}},\ }\bibfield  {title} {\bibinfo {title} {{\textit{Fractional Quantum Anomalous Hall States in Twisted Bilayer MoTe$_2$ and WSe$_2$}}},\ }\href@noop {} {\bibfield  {journal} {\bibinfo  {journal} {Phys. Rev. B}\ }\textbf {\bibinfo {volume} {108}},\ \bibinfo {pages} {085117} (\bibinfo {year} {2023})}\BibitemShut {NoStop}%
\bibitem [{\citenamefont {Wang}\ \emph {et~al.}(2024)\citenamefont {Wang}, \citenamefont {Zhang}, \citenamefont {Liu}, \citenamefont {He}, \citenamefont {Xu}, \citenamefont {Ran}, \citenamefont {Cao},\ and\ \citenamefont {Xiao}}]{wang2024fractional}%
  \BibitemOpen
  \bibfield  {author} {\bibinfo {author} {\bibfnamefont {C.}~\bibnamefont {Wang}}, \bibinfo {author} {\bibfnamefont {X.-W.}\ \bibnamefont {Zhang}}, \bibinfo {author} {\bibfnamefont {X.}~\bibnamefont {Liu}}, \bibinfo {author} {\bibfnamefont {Y.}~\bibnamefont {He}}, \bibinfo {author} {\bibfnamefont {X.}~\bibnamefont {Xu}}, \bibinfo {author} {\bibfnamefont {Y.}~\bibnamefont {Ran}}, \bibinfo {author} {\bibfnamefont {T.}~\bibnamefont {Cao}},\ and\ \bibinfo {author} {\bibfnamefont {D.}~\bibnamefont {Xiao}},\ }\bibfield  {title} {\bibinfo {title} {{\textit{Fractional Chern Insulator in Twisted Bilayer MoTe$_2$}}},\ }\href@noop {} {\bibfield  {journal} {\bibinfo  {journal} {Phys. Rev. Lett.}\ }\textbf {\bibinfo {volume} {132}},\ \bibinfo {pages} {036501} (\bibinfo {year} {2024})}\BibitemShut {NoStop}%
\bibitem [{\citenamefont {Sharma}\ \emph {et~al.}(2024)\citenamefont {Sharma}, \citenamefont {Peng},\ and\ \citenamefont {Sheng}}]{sharma2024topological}%
  \BibitemOpen
  \bibfield  {author} {\bibinfo {author} {\bibfnamefont {P.}~\bibnamefont {Sharma}}, \bibinfo {author} {\bibfnamefont {Y.}~\bibnamefont {Peng}},\ and\ \bibinfo {author} {\bibfnamefont {D.}~\bibnamefont {Sheng}},\ }\bibfield  {title} {\bibinfo {title} {{\textit{Topological Quantum Phase Transitions Driven by a Displacement Field in Twisted MoTe$_2$ Bilayers}}},\ }\href@noop {} {\bibfield  {journal} {\bibinfo  {journal} {Phys. Rev. B}\ }\textbf {\bibinfo {volume} {110}},\ \bibinfo {pages} {125142} (\bibinfo {year} {2024})}\BibitemShut {NoStop}%
\bibitem [{\citenamefont {Saminadayar}\ \emph {et~al.}(1997)\citenamefont {Saminadayar}, \citenamefont {Glattli}, \citenamefont {Jin},\ and\ \citenamefont {Etienne}}]{saminadayar1997observation}%
  \BibitemOpen
  \bibfield  {author} {\bibinfo {author} {\bibfnamefont {L.}~\bibnamefont {Saminadayar}}, \bibinfo {author} {\bibfnamefont {D.}~\bibnamefont {Glattli}}, \bibinfo {author} {\bibfnamefont {Y.}~\bibnamefont {Jin}},\ and\ \bibinfo {author} {\bibfnamefont {B.}~\bibnamefont {Etienne}},\ }\bibfield  {title} {\bibinfo {title} {{\textit{Observation of the e/3 Fractionally Charged Laughlin Quasiparticle}}},\ }\href@noop {} {\bibfield  {journal} {\bibinfo  {journal} {Phys. Rev. Lett.}\ }\textbf {\bibinfo {volume} {79}},\ \bibinfo {pages} {2526} (\bibinfo {year} {1997})}\BibitemShut {NoStop}%
\bibitem [{\citenamefont {Zhou}\ \emph {et~al.}(2019)\citenamefont {Zhou}, \citenamefont {Chen}, \citenamefont {Liu}, \citenamefont {Sochnikov}, \citenamefont {Bollinger}, \citenamefont {Han}, \citenamefont {Zhu}, \citenamefont {He}, \citenamefont {Bozovi{\'c}},\ and\ \citenamefont {Natelson}}]{zhou2019electron}%
  \BibitemOpen
  \bibfield  {author} {\bibinfo {author} {\bibfnamefont {P.}~\bibnamefont {Zhou}}, \bibinfo {author} {\bibfnamefont {L.}~\bibnamefont {Chen}}, \bibinfo {author} {\bibfnamefont {Y.}~\bibnamefont {Liu}}, \bibinfo {author} {\bibfnamefont {I.}~\bibnamefont {Sochnikov}}, \bibinfo {author} {\bibfnamefont {A.~T.}\ \bibnamefont {Bollinger}}, \bibinfo {author} {\bibfnamefont {M.-G.}\ \bibnamefont {Han}}, \bibinfo {author} {\bibfnamefont {Y.}~\bibnamefont {Zhu}}, \bibinfo {author} {\bibfnamefont {X.}~\bibnamefont {He}}, \bibinfo {author} {\bibfnamefont {I.}~\bibnamefont {Bozovi{\'c}}},\ and\ \bibinfo {author} {\bibfnamefont {D.}~\bibnamefont {Natelson}},\ }\bibfield  {title} {\bibinfo {title} {{\textit{Electron pairing in the pseudogap state revealed by shot noise in copper oxide junctions}}},\ }\href@noop {} {\bibfield  {journal} {\bibinfo  {journal} {Nature}\ }\textbf {\bibinfo {volume} {572}},\ \bibinfo {pages} {493} (\bibinfo {year} {2019})}\BibitemShut {NoStop}%
\bibitem [{\citenamefont {Chen}\ \emph {et~al.}(2023)\citenamefont {Chen}, \citenamefont {Lowder}, \citenamefont {Bakali}, \citenamefont {Andrews}, \citenamefont {Schrenk}, \citenamefont {Waas}, \citenamefont {Svagera}, \citenamefont {Eguchi}, \citenamefont {Prochaska}, \citenamefont {Wang}, \citenamefont {Setty}, \citenamefont {Sur}, \citenamefont {Si}, \citenamefont {Paschen},\ and\ \citenamefont {Natelson}}]{chen2023shot}%
  \BibitemOpen
  \bibfield  {author} {\bibinfo {author} {\bibfnamefont {L.}~\bibnamefont {Chen}}, \bibinfo {author} {\bibfnamefont {D.~T.}\ \bibnamefont {Lowder}}, \bibinfo {author} {\bibfnamefont {E.}~\bibnamefont {Bakali}}, \bibinfo {author} {\bibfnamefont {A.~M.}\ \bibnamefont {Andrews}}, \bibinfo {author} {\bibfnamefont {W.}~\bibnamefont {Schrenk}}, \bibinfo {author} {\bibfnamefont {M.}~\bibnamefont {Waas}}, \bibinfo {author} {\bibfnamefont {R.}~\bibnamefont {Svagera}}, \bibinfo {author} {\bibfnamefont {G.}~\bibnamefont {Eguchi}}, \bibinfo {author} {\bibfnamefont {L.}~\bibnamefont {Prochaska}}, \bibinfo {author} {\bibfnamefont {Y.}~\bibnamefont {Wang}}, \bibinfo {author} {\bibfnamefont {C.}~\bibnamefont {Setty}}, \bibinfo {author} {\bibfnamefont {S.}~\bibnamefont {Sur}}, \bibinfo {author} {\bibfnamefont {Q.}~\bibnamefont {Si}}, \bibinfo {author} {\bibfnamefont {S.}~\bibnamefont {Paschen}},\ and\ \bibinfo {author} {\bibfnamefont {D.}~\bibnamefont {Natelson}},\ }\bibfield  {title} {\bibinfo {title} {{\textit{Shot Noise in
  a Strange Metal}}},\ }\href@noop {} {\bibfield  {journal} {\bibinfo  {journal} {Science}\ }\textbf {\bibinfo {volume} {382}},\ \bibinfo {pages} {907} (\bibinfo {year} {2023})}\BibitemShut {NoStop}%
\bibitem [{\citenamefont {Niu}\ \emph {et~al.}(2024)\citenamefont {Niu}, \citenamefont {Bastiaans}, \citenamefont {Ge}, \citenamefont {Tomar}, \citenamefont {Jesudasan}, \citenamefont {Raychaudhuri}, \citenamefont {Karrer}, \citenamefont {Kleiner}, \citenamefont {Koelle}, \citenamefont {Barbier} \emph {et~al.}}]{niu2024shot}%
  \BibitemOpen
  \bibfield  {author} {\bibinfo {author} {\bibfnamefont {J.}~\bibnamefont {Niu}}, \bibinfo {author} {\bibfnamefont {K.~M.}\ \bibnamefont {Bastiaans}}, \bibinfo {author} {\bibfnamefont {J.-F.}\ \bibnamefont {Ge}}, \bibinfo {author} {\bibfnamefont {R.}~\bibnamefont {Tomar}}, \bibinfo {author} {\bibfnamefont {J.}~\bibnamefont {Jesudasan}}, \bibinfo {author} {\bibfnamefont {P.}~\bibnamefont {Raychaudhuri}}, \bibinfo {author} {\bibfnamefont {M.}~\bibnamefont {Karrer}}, \bibinfo {author} {\bibfnamefont {R.}~\bibnamefont {Kleiner}}, \bibinfo {author} {\bibfnamefont {D.}~\bibnamefont {Koelle}}, \bibinfo {author} {\bibfnamefont {A.}~\bibnamefont {Barbier}}, \emph {et~al.},\ }\bibfield  {title} {\bibinfo {title} {{\textit{Why shot noise does not generally detect pairing in mesoscopic superconducting tunnel junctions}}},\ }\href@noop {} {\bibfield  {journal} {\bibinfo  {journal} {Phys. Rev. Lett.}\ }\textbf {\bibinfo {volume} {132}},\ \bibinfo {pages} {076001} (\bibinfo {year} {2024})}\BibitemShut {NoStop}%
\bibitem [{\citenamefont {Blankenbecler}\ \emph {et~al.}(1981)\citenamefont {Blankenbecler}, \citenamefont {Scalapino},\ and\ \citenamefont {Sugar}}]{blankenbecler1981monte}%
  \BibitemOpen
  \bibfield  {author} {\bibinfo {author} {\bibfnamefont {R.}~\bibnamefont {Blankenbecler}}, \bibinfo {author} {\bibfnamefont {D.~J.}\ \bibnamefont {Scalapino}},\ and\ \bibinfo {author} {\bibfnamefont {R.~L.}\ \bibnamefont {Sugar}},\ }\bibfield  {title} {\bibinfo {title} {{\textit{Monte Carlo Calculations of Coupled Boson-Fermion Systems. {I}}}},\ }\href@noop {} {\bibfield  {journal} {\bibinfo  {journal} {Phys. Rev. D}\ }\textbf {\bibinfo {volume} {24}},\ \bibinfo {pages} {2278} (\bibinfo {year} {1981})}\BibitemShut {NoStop}%
\bibitem [{\citenamefont {Assaad}\ and\ \citenamefont {Evertz}(2008)}]{assaad2008world}%
  \BibitemOpen
  \bibfield  {author} {\bibinfo {author} {\bibfnamefont {F.~F.}\ \bibnamefont {Assaad}}\ and\ \bibinfo {author} {\bibfnamefont {H.~G.}\ \bibnamefont {Evertz}},\ }\bibfield  {title} {\bibinfo {title} {{\textit{World-Line and Determinantal Quantum Monte Carlo Methods for Spins, Phonons and Electrons}}},\ }in\ \href@noop {} {\emph {\bibinfo {booktitle} {Computational Many-Particle Physics}}},\ \bibinfo {series} {Lecture Notes in Physics}, Vol.\ \bibinfo {volume} {739},\ \bibinfo {editor} {edited by\ \bibinfo {editor} {\bibfnamefont {H.}~\bibnamefont {Fehske}}, \bibinfo {editor} {\bibfnamefont {R.}~\bibnamefont {Schneider}},\ and\ \bibinfo {editor} {\bibfnamefont {A.}~\bibnamefont {Wei{\ss}e}}}\ (\bibinfo  {publisher} {Springer},\ \bibinfo {address} {Berlin, Heidelberg},\ \bibinfo {year} {2008})\ p.\ \bibinfo {pages} {277}\BibitemShut {NoStop}%
\bibitem [{\citenamefont {Bistritzer}\ and\ \citenamefont {MacDonald}(2011)}]{bistritzer2011moire}%
  \BibitemOpen
  \bibfield  {author} {\bibinfo {author} {\bibfnamefont {R.}~\bibnamefont {Bistritzer}}\ and\ \bibinfo {author} {\bibfnamefont {A.~H.}\ \bibnamefont {MacDonald}},\ }\bibfield  {title} {\bibinfo {title} {Moir{\'e} bands in twisted double-layer graphene},\ }\href@noop {} {\bibfield  {journal} {\bibinfo  {journal} {Proceedings of the National Academy of Sciences}\ }\textbf {\bibinfo {volume} {108}},\ \bibinfo {pages} {12233} (\bibinfo {year} {2011})}\BibitemShut {NoStop}%
\bibitem [{\citenamefont {Koshino}\ \emph {et~al.}(2018)\citenamefont {Koshino}, \citenamefont {Yuan}, \citenamefont {Koretsune}, \citenamefont {Ochi}, \citenamefont {Kuroki},\ and\ \citenamefont {Fu}}]{koshino2018maximally}%
  \BibitemOpen
  \bibfield  {author} {\bibinfo {author} {\bibfnamefont {M.}~\bibnamefont {Koshino}}, \bibinfo {author} {\bibfnamefont {N.~F.~Q.}\ \bibnamefont {Yuan}}, \bibinfo {author} {\bibfnamefont {T.}~\bibnamefont {Koretsune}}, \bibinfo {author} {\bibfnamefont {M.}~\bibnamefont {Ochi}}, \bibinfo {author} {\bibfnamefont {K.}~\bibnamefont {Kuroki}},\ and\ \bibinfo {author} {\bibfnamefont {L.}~\bibnamefont {Fu}},\ }\bibfield  {title} {\bibinfo {title} {{\textit{Maximally Localized Wannier Orbitals and the Extended Hubbard Model for Twisted Bilayer Graphene}}},\ }\href@noop {} {\bibfield  {journal} {\bibinfo  {journal} {Phys. Rev. X}\ }\textbf {\bibinfo {volume} {8}},\ \bibinfo {pages} {031087} (\bibinfo {year} {2018})}\BibitemShut {NoStop}%
\bibitem [{\citenamefont {Xiao}\ \emph {et~al.}(2012)\citenamefont {Xiao}, \citenamefont {Liu}, \citenamefont {Feng}, \citenamefont {Xu},\ and\ \citenamefont {Yao}}]{xiao2012coupled}%
  \BibitemOpen
  \bibfield  {author} {\bibinfo {author} {\bibfnamefont {D.}~\bibnamefont {Xiao}}, \bibinfo {author} {\bibfnamefont {G.-B.}\ \bibnamefont {Liu}}, \bibinfo {author} {\bibfnamefont {W.}~\bibnamefont {Feng}}, \bibinfo {author} {\bibfnamefont {X.}~\bibnamefont {Xu}},\ and\ \bibinfo {author} {\bibfnamefont {W.}~\bibnamefont {Yao}},\ }\bibfield  {title} {\bibinfo {title} {\textit{Coupled spin and valley physics in monolayers of MoS$_2$ and other group-VI dichalcogenides}},\ }\href@noop {} {\bibfield  {journal} {\bibinfo  {journal} {Phys. Rev. Lett.}\ }\textbf {\bibinfo {volume} {108}},\ \bibinfo {pages} {196802} (\bibinfo {year} {2012})}\BibitemShut {NoStop}%
\bibitem [{\citenamefont {Wu}\ \emph {et~al.}(2019)\citenamefont {Wu}, \citenamefont {Lovorn}, \citenamefont {Tutuc}, \citenamefont {Martin},\ and\ \citenamefont {MacDonald}}]{wu2019topological}%
  \BibitemOpen
  \bibfield  {author} {\bibinfo {author} {\bibfnamefont {F.}~\bibnamefont {Wu}}, \bibinfo {author} {\bibfnamefont {T.}~\bibnamefont {Lovorn}}, \bibinfo {author} {\bibfnamefont {E.}~\bibnamefont {Tutuc}}, \bibinfo {author} {\bibfnamefont {I.}~\bibnamefont {Martin}},\ and\ \bibinfo {author} {\bibfnamefont {A.}~\bibnamefont {MacDonald}},\ }\bibfield  {title} {\bibinfo {title} {\textit{Topological insulators in twisted transition metal dichalcogenide homobilayers}},\ }\href@noop {} {\bibfield  {journal} {\bibinfo  {journal} {Phys. Rev. Lett.}\ }\textbf {\bibinfo {volume} {122}},\ \bibinfo {pages} {086402} (\bibinfo {year} {2019})}\BibitemShut {NoStop}%
\end{thebibliography}%

\section*{Acknowledgments}
The authors acknowledge Mark Dean, Ben Feldman, Kaden Hazzard, Yu He, Matteo Mitrano, Douglas Natelson, Yinming Shao, and Yonglong Xie for insightful discussions. This work is primarily supported by the U.S. Department of Energy, Office of Science, Basic Energy Sciences, under Early Career Award No.~DE-SC0024524. D.S. also acknowledges the Scialog Grant No. SA-QMI-2025-081b through Research Corporation for Science Advancement. The simulations presented in this work used resources of the National Energy Research Scientific Computing Center, a U.S. Department of Energy Office of Science User Facility located at Lawrence Berkeley National Laboratory, operated under Contract No.~DE-AC02-05CH11231.

\section{Author contributions}
Y.W. conceived the idea of the project. S.D. performed theoretical derivations and DQMC simulations. P.S. performed DMRG and ED simulations. Z.S., J.-X. L., and S.U. analyzed the response theory and experimental validity. S.D., P.S., and Y.W. wrote the manuscript with inputs from all other authors. 

\section{Competing interests}
The authors declare no competing interests.

\end{document}